\documentclass[aps,amssymb,amsmath,reprint]{revtex4-1} 
\usepackage{graphicx}  
\usepackage{dcolumn}   
\usepackage{physics}
\usepackage{color}
\usepackage{appendix}

\definecolor{orange}{rgb}{0.8, 0.3, 0}

\definecolor{blueviolet}{rgb}{0.2, 0.2, 0.6}
\usepackage[pdftex, 
bookmarks=true, 
colorlinks=true,
allcolors=blueviolet,
pdfstartview={FitH}, 
]{hyperref}

\newcommand{\geff}{\gamma_\text{eff}}

\begin{document}

\title{Spatially-resolved decoherence of donor spins in silicon strained by a metallic electrode}

\author{V.~Ranjan$^{1,6}$}

\author{B.~Albanese$^{1}$}

\author{E.~Albertinale$^{1}$}

\author{E.~Billaud$^{1}$}

\author{D.~Flanigan$^{1}$}

\author{J.~J.~Pla$^{2}$}

\author{T.~Schenkel$^{3}$}

\author{D.~Vion$^{1}$}

\author{D.~Esteve$^{1}$}

\author{E.~Flurin$^{1}$}

\author{J.~J.~L.~Morton$^{4}$}

\author{Y.~M.~Niquet$^{5}$}

\author{P.~Bertet$^{1}$}
\email{patrice.bertet@cea.fr}

\affiliation{$^{1}$Universit\'e Paris-Saclay, CEA, CNRS, SPEC, 91191 Gif-sur-Yvette Cedex, France}

\affiliation{$^{2}$School of Electrical Engineering and Telecommunications, University of New South Wales,
Anzac Parade, Sydney, New South Wales 2052, Australia}

\affiliation{$^{3}$Accelerator Technology and Applied Physics Division, Lawrence Berkeley National Laboratory, Berkeley, California 94720, USA}

\affiliation{$^{4}$London Centre for Nanotechnology, University College London, London WC1H 0AH, United Kingdom}

\affiliation{$^{5}$Universit\'e Grenoble Alpes, CEA, IRIG-MEM-L\_SIM, 38000 Grenoble, France}

\affiliation{$^6$National Physical Laboratory, Teddington TW11 0LW, United Kingdom }

\begin{abstract}
Electron spins are amongst the most coherent solid-state systems known, however, to be used in devices for quantum sensing and information processing applications, they must be typically placed near interfaces. Understanding and mitigating the impacts of such interfaces on the coherence and spectral properties of electron spins is critical to realize such applications, but is also challenging: inferring such data from single-spin studies requires many measurements to obtain meaningful results, while ensemble measurements typically give averaged results that hide critical information. Here, we report a comprehensive study of the coherence of near-surface bismuth donor spins in 28-silicon at millikelvin temperatures. In particular, we use strain-induced frequency shifts caused by a metallic electrode to make spatial maps of spin coherence as a function of depth and position relative to the electrode. By measuring magnetic-field-insensitive clock transitions we separate magnetic noise caused by surface spins from charge noise. Our results include quantitative models of the strain-split spin resonance spectra and extraction of paramagnetic impurity concentrations at the silicon surface. The interplay of these decoherence mechanisms for such near-surface electron spins is critical for their application in quantum technologies, while the combination of the strain splitting and clock transition extends the coherence lifetimes by up to two orders of magnitude, reaching up to 300 ms at a mean depth of only 100 nm.
The technique we introduce here to spatially map coherence in near-surface ensembles is directly applicable to other spin systems of active interest, such as defects in diamond, silicon carbide, and rare earth ions in optical crystals.

\end{abstract}

\maketitle

\section{Introduction}
The electron and nuclear spins of donors in silicon are promising qubit candidates for solid-state quantum computing~\cite{kane_silicon-based_1998,morello_single-shot_2010,pla_single-atom_2012,veldhorst_addressable_2014,muhonen_storing_2014,yang_operation_2020,petit_universal_2020,morello_donor_nodate}. This is due to their exceptionally long coherence times when the silicon susbtrate is isotopically enriched in the nuclear-spin-free isotope $^{28}$Si, reaching seconds for the donor electron spin ~\cite{tyryshkin_electron_2012,wolfowicz_atomic_2013} and minutes for its nuclear spin~\cite{muhonen_storing_2014}. Donors in silicon are therefore also suitable for implementing long-lived and multi-mode microwave quantum memories for superconducting qubits~\cite{julsgaard_quantum_2013,morton_storing_2018,ranjan_multimode_2020-1}.

For quantum information processing applications, the donors need to be close to the silicon/silicon-oxide interface and also to the metallic electrodes used to apply the electromagnetic signals enabling quantum state manipulation and readout. Interfaces, however, have large defect densities, consisting of dangling bonds, surface adsorbents, and tunneling two-level-systems (TLS). Moreover, the electrodes are often made with materials that may have thermal expansion coefficients different from the substrate, resulting in spatially-dependent mechanical strain. For future donor-based quantum devices, it is essential to understand how the spectrum and coherence time of donors in silicon are modified by the proximity of interfaces with the surface oxide~\cite{schenkel_electrical_2006,de_sousa_dangling-bond_2007,paik_t_1_2010} and with the electrodes. While single-donor devices provide detailed information on the behavior of individual dopants ~\cite{muhonen_storing_2014,tenberg_electron_2019}, they would require a prohibitively large number of measurements for extracting spatially-resolved properties on a larger scale. Here, we achieve this goal with an ensemble of donors in silicon in the vicinity of a micron-wide aluminum electrode. Using strain-induced frequency shifts combined with extensive coherence time measurements, we infer magnetic and charge noise as a function of depth and position relative to the electrode.

The measurements are made possible by the properties of bismuth donors in silicon, a particularly interesting system to probe the physical mechanisms at play near interfaces. Due to the strong hyperfine interaction of the donor electron spin $S=1/2$ with the $I=9/2$ nuclear spin of the bismuth atom, bismuth donors are sensitive to strain~\cite{mansir_linear_2018} and electric fields, and have transitions with widely varying effective gyromagnetic ratios. This allows us to distinguish the various sources of noise by comparing coherence times on different transitions.

Our device consists of an ensemble of shallow-implanted bismuth donors in a silicon substrate isotopically-enriched in $^{28}\mathrm{Si}$, magnetically coupled to an aluminum wire deposited on top of the substrate. The wire is the inductor in a superconducting $LC$ resonator used for quantum-limited Electron Paramagnetic Resonance (EPR) spectroscopy at millikelvin temperatures~\cite{bienfait_reaching_2016,probst_inductive-detection_2017,ranjan_electron_2020} (Sections~\ref{sec:Bidopants} and \ref{sec:methods}). The bismuth donor spin spectrum of a similar device was shown to be governed by the strain imparted by the differential thermal contraction of the aluminum wire with respect to the underlying silicon substrate~\cite{pla_strain-induced_2018,osullivan_spin-resonance_2020}. Here we confirm these results by showing that such a model also quantitatively predicts the lineshape when varying the wire width (Section~\ref{sec:sepctroscopy}). We furthermore use the phenomenon of instantaneous diffusion to demonstrate that strain is locally inhomogeneous (Section~\ref{sec:ID}). We then study the Hahn-echo coherence time $T_2$ for various spin transitions, wire geometries, and strain values. We observe a strong dependence of $T_2$ on strain. Because in our device geometry the strained donors are also closer to the surface, this provides evidence that $T_2$ is limited by magnetic noise originating from the device interfaces, in agreement with previous work~\cite{schenkel_electrical_2006,paik_t_1_2010}. Taking advantage of the the strain gradient around the wire, our measurements provide some amount of spatial resolution, indicating a paramagnetic defect density of $4\times 10^{12}~\mathrm{cm}^{-2}$ away from the wire, and $ 10^{12}~\mathrm{cm}^{-2}$ below the wire (Section~\ref{sec:magneticNoise}). We finally measure $T_2$ on a magnetic-noise-insensitive Clock Transition, finding again a strong dependence on strain, with the largest measured value being $T_2 = 300$\,ms. We argue that this $T_2$ is limited by charge noise at the silicon/silicon oxide interface (Section~\ref{sec:non-magneticnoise}). We support our findings with complementary measurements of both the magnetic and charge noise dependence on pulse sequence (via dynamical decoupling experiments), temperature, and artificially introduced microwave noise (Section~\ref{sec:DDandTemp}).

\begin{figure*}[t!]
  \includegraphics[width=2\columnwidth]{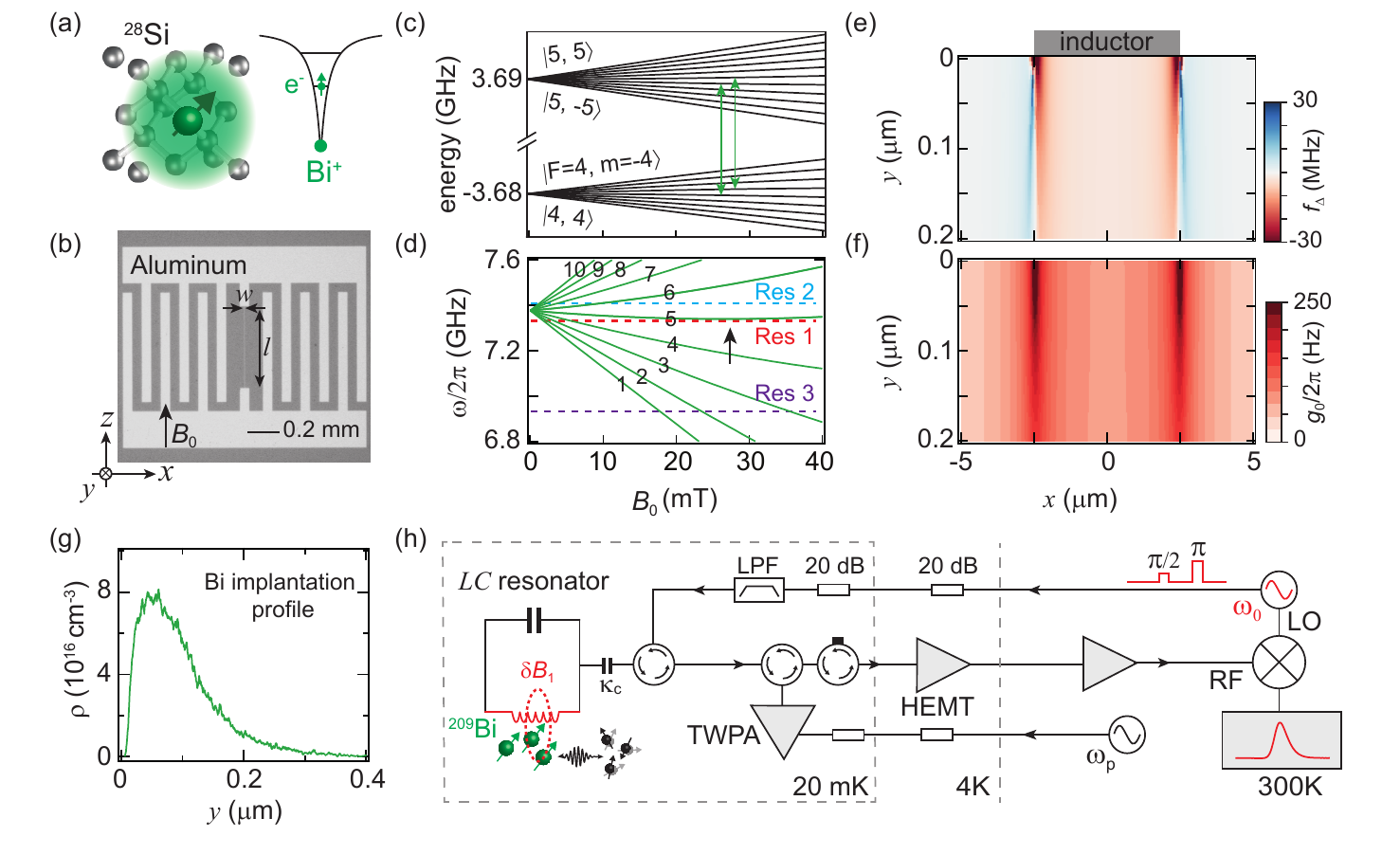}
  \caption{\label{fig:strainedbismuth}
 Bismuth spins in silicon coupled to $LC$ resonators. (a) A schematic of a bismuth atom and associated electron cloud in the silicon lattice. The donor electron stays bound to the Bi$^+$ core at low temperatures due to Coulomb potential. (b) Optical picture of one of the 50~nm thick aluminum $LC$ resonators directly fabricated on the Si substrate ($xz$ plane and $y=0$). The inductor is of width $w$ and length $l$, while inter-digitated finger capacitors are $50~\mu$m wide and apart. (c) Numerically calculated eigenenergies $|F,~m \rangle$ and (d) EPR allowed transition frequencies of Si:Bi at low magnetic fields. Transitions are numbered for referencing. A clock transition near 27~mT is marked by the arrow. (e) Strain induced Larmor frequency shift $f_\Delta$ and (f) spin-photon coupling strength $g_0$ for the first transition in Res1. (g) The implantation profile $\rho (y)$ of the Bi dopants. (h) The measurement setup at different temperature stages. Input pulses are sent through a series of attenuators and low pass filters (LPF). The reflected output signals are routed via a circulator and amplified with a Josephson traveling wave parametric amplifier (TWPA) before further amplification and demodulation at room temperature. The shallow-implanted Bi spins interact with fluctuating impurity spins on the substrate surface, and are detected via inductive coupling to the resonator. }
\end{figure*}

\section{Bismuth dopants in silicon}
\label{sec:Bidopants}

We briefly summarize the properties of bismuth donor spins in silicon that are relevant to this work. More details can be found in Ref.~\onlinecite{feher_electron_1959,mohammady_bismuth_2010} and  Appendix~\ref{app:Sx}.

As a group-V element, bismuth is an electron donor when inserted in the silicon lattice where it forms four covalent bonds. At low temperature, the extra electron ($S= 1/2$) can be trapped around the Bi$^+$ ion (with nuclear spin $I = 9/2$) by the Coulomb interaction, forming a coupled electron-nuclear spin system~\cite{feher_electron_1959}. Group-V donors have a symmetric $s$-wave electronic wavefunction in the ground state, similar to a hydrogen atom, albeit with a Bohr radius spanning several lattice constants~[Fig.~\ref{fig:strainedbismuth}(a)]. The Hamiltonian of a donor spin system under a static magnetic field $B_0$ is given by

\begin{equation}
\label{eq:ham}
  H/\hbar = A \bold{S} \cdot \bold{I} + (\gamma_e \bold{S}+ \gamma_n \bold{I})  \cdot \bold{B_0},  
\end{equation}
where $\gamma_e/2\pi = 28~$MHz/mT and $\gamma_n / 2\pi = 7~$kHz/mT are the electronic and nuclear gyromagnetic ratios, and $A/2\pi = 1.475$\,GHz is the hyperfine coupling constant between electron and nuclear spins. One specificity of bismuth donor spins compared to other group-V donors such as phosphorus is the large value of $A$. This makes it possible to explore the low-field limit $\gamma_e |B_0| \ll A$ (corresponding to $B_0 < 10-20\, \mathrm{mT}$), in which the energy eigenstates wavefunctions are hybridized electro-nuclear spin wavefunctions dominated by the hyperfine interaction. All the results presented here are in this low-field limit.

Under Hamiltonian Eq.\ref{eq:ham}, the total angular momentum projection $m$ on the $z-$direction along which $B_0$ is applied is a good quantum number, and the total angular momentum (of the operator $\mathbf{F} = \mathbf{S}+\mathbf{I}$) an approximate one in the low-field limit. The energy levels can be determined numerically or analytically~\cite{mohammady_bismuth_2010} and are shown in Fig.~\ref{fig:strainedbismuth}(c). They can be approximated by states $|F,m \rangle$, and are distributed into a low-energy $(F = 4)$ manifold of 9 states and a high-energy $(F = 5)$ manifold of 11 states separated by a zero field splitting (ZFS) of $5A/2\pi = 7.375~$GHz. The ZFS makes it particularly suitable for EPR spectroscopy using superconducting resonators, since it can be performed even at low magnetic fields if the resonator frequency is close to the ZFS. 

EPR transitions driven resonantly by a transverse ac magnetic field $B_1$ are allowed between levels satisfying $|\Delta m| = 1$. In Fig.~\ref{fig:strainedbismuth}(d) are shown the $18$ allowed EPR transition frequencies as a function of $B_0$ in the vicinity of the ZFS. Because $8$ pairs of these transitions are quasi-degenerate, only $10$ curves are resolved. It can be shown that in the low-field limit (except near clock transition), the transition frequency $\omega(m)$ of $|4,m\rangle \leftrightarrow |5,m+1\rangle$ with $m \in [-4,~4]$ as well as $|4,m+1\rangle \leftrightarrow |5,m \rangle$ with $m \in [-5,3]$ is given by

\begin{equation}
\label{eq:lowfield}
    \omega(m)\approx 5 A + \frac{2m + 1}{10}\gamma_e B_0. 
\end{equation}
We will refer to these $10$ resonances by labels $i=1,2,...,10$ from the lowest to the highest frequency [see Fig.~\ref{fig:strainedbismuth}(d)]. It is notable that the effective gyromagnetic ratio $\gamma_\text{eff} \equiv \text{d}\omega/\text{d}B_0 \simeq (2m+1) \gamma_e/10$ strongly depends on $m$ and therefore on $i$. In the following, we exploit the transition-dependent $\gamma_\text{eff}$ and hence variable sensitivity to magnetic noise in our analysis of decoherence. 

In addition to the obvious sensitivity to magnetic fields via the Zeeman effect, bismuth donor spins are also sensitive to strain (described by the tensor $\mathbf{\varepsilon}$) and electric field $\mathbf{E}$ via the hyperfine coupling constant $A(\mathbf{\varepsilon},\mathbf{E})$~\cite{mansir_linear_2018}. 

Strain modifies the donor wavefunction, and therefore also the Fermi-contact hyperfine constant $A$. This effect was quantitatively studied in Ref.~\onlinecite{mansir_linear_2018}, where the change $\Delta A(\varepsilon)$ with respect to the unperturbed value $A_0$ was found to be

\begin{multline}
\label{eq:fDeltastrain}
      \frac{\Delta A(\mathbf{\varepsilon})}{A_0} = \frac{K}{3} (\epsilon_{xx} + \epsilon_{yy} +\epsilon_{zz})  \\ 
      - \frac{L}{2} \left[ (\epsilon_{xx} - \epsilon_{yy})^2 +
    (\epsilon_{yy} - \epsilon_{zz})^2 + (\epsilon_{zz} - \epsilon_{xx})^2 \right],
\end{multline}

\noindent where $\epsilon_{ii}$ are the uniaxial components of the strain tensor, $K=19.1$ is determined by the measurements, and $L=9720$ by the valley repopulation model (VRM)~\cite{mansir_linear_2018}. For strain values $< 10^{-3}$, the hydrostatic component of strain dominates the shift in $\omega(m)$ whose sign, unlike the prediction from the VRM, can be both positive and negative. Besides modifying the spin Hamiltonian parameters, strain can also affect the spin-lattice relaxation time as shown recently~\cite{tenberg_electron_2019}.

The hyperfine coupling constant $A$ is also dependent on an applied electric field $\mathbf{E}$. This Stark effect $\Delta A(\mathbf{E})/A_0 = \eta E^2$ is quadratic with $\eta = (-0.26 \pm 0.05) \times 10^{-3}~\mu \mathrm{m}^2 / \mathrm{V}^2$ for bismuth donors, a value obtained by measurements and close to the value calculated by a multivalley effective mass theory~\cite{pica_hyperfine_2014}. 

Interestingly, because each transition frequency satisfies $\partial \omega / \partial A \simeq  5$ (see Eq.~\ref{eq:lowfield}) in the low-field regime, both strain and electric field have a very simple effect on the complex bismuth energy spectrum: they cause a frequency shift $f_\Delta (\mathbf{\varepsilon},\mathbf{E}) = 5 [\Delta A (\mathbf{\varepsilon}) + \Delta A (\mathbf{E})] / 2\pi$ identical for all bismuth donor transitions.

Another relevant feature of bismuth donor spins to this work is the existence of biasing conditions where $\gamma_\text{eff} = 0$. Such Clock Transitions (CT)~\cite{wolfowicz_atomic_2013} are the result of the competition between the hyperfine and Zeeman interaction, in a regime where the low-field approximation Eq.~\ref{eq:lowfield} is no longer strictly valid. At a CT, the donor becomes insensitive to magnetic noise to first order, allowing us to probe other sources of noise. In the following we will use the CT that occurs at $27$\,mT [see Fig.~\ref{fig:strainedbismuth}(d)] to investigate non-magnetic sources of decoherence, such as charge noise which is known to exist in silicon CMOS structures.

\section{Materials and Methods}
\label{sec:methods}
Our device is fabricated on a silicon (100) chip having an isotopically purified epitaxial layer of $^{28}$Si (0.05\%  of $^{29}\text{Si}$), in which bismuth atoms are implanted about $75$~nm below the surface [Fig.~\ref{fig:strainedbismuth}(e)]. The chip is cleaned in a Piranha solution (3:1 mix of $\text{H}_2\text{SO}_4$ and $\text{H}_2 \text{O}_2$ at $120^\circ$C for 15 minutes). The chip has thus a native silicon oxide layer. Finally, three LC superconducting resonators are patterned on top by standard electron-beam lithography, followed by oxygen ashing and evaporation of aluminum. The inductors of the resonators are fabricated along the $[011]$ crystallographic axis ($z$-frame axis), and have different widths ($w = 5~\mu$m, 2~$\mu$m and 1~$\mu$m for Res1, Res2 and Res3 respectively). The resonance frequencies $\omega_0$ are designed to be close to the ZFS for accessing multiple transitions at small magnetic fields and thus preserving low loss rates of the superconducting resonators.

The chip is mounted inside a copper box with a hole and an antenna terminating a $50~\Omega$ line. The line and the antenna fixes the energy decay rates $\kappa_c$ of the resonators, and allow to drive and measure the spins in reflection (see Fig.~\ref{fig:strainedbismuth}(e) and Ref.~\onlinecite{bienfait_reaching_2016}). The copper box is installed inside a home-built coil that produces a magnetic field parallel to the inductor. All measurements are performed in a dilution refrigerator at $\sim 15$~mK (unless mentioned otherwise), its base temperature. By measuring the reflection coefficient of each resonator with a vector network analyzer, we extract $\kappa_c$, the internal loss rate $\kappa_i$, and the total loss rate $\kappa = \kappa_i + \kappa_c$, at an average intra-cavity photon number $n_\text{cav} \sim 1$ (see Table~\ref{table1}). All spectroscopy and coherence time measurements are achieved using a two-pulse Hahn echo sequence $\pi/2 -\tau - \pi - \tau - echo$, and computing the echo integral $\chi_e$. Control pulses are either square, or shaped~\cite{probst_shaped_2019} to exceed the resonator bandwidth limitation. Their absolute amplitude at the level of the resonators is deduced from the Rabi oscillations measurements (Appendix~\ref{app:Sx}).

Bismuth spins are inductively coupled to the resonator with a single spin photon coupling strength, $g_0 = \gamma_e \langle 4,m| S_x |5, m\pm 1 \rangle \cdot \delta B_1$ where $\delta B_1$ is the rms vacuum fluctuations of the resonator magnetic field~\cite{bienfait_reaching_2016}. The corresponding Rabi frequency is given by is given by $\Omega_\text{R} = 2 g_0 \sqrt{n_\text{cav}}$. We calculate $\delta B_1$ for different resonators using finite element simulations by solving for a current $\delta i = \omega_0 \sqrt{\hbar/2Z_0}$ flowing through the inductor with an appropriate spatial density in the superconducting film~\cite{duzer_principles_1999}. Here, $Z_0$ is the impedance of the resonator estimated independently from electromagnetic simulations.  

One specificity of our setup is the small mode volume of the superconducting resonators, enabling $g_0/2\pi$ to reach large values $10^2-10^3~\text{Hz}$. This enhances the spin radiative energy relaxation rate via the Purcell effect~\cite{purcell_spontaneous_1946,bienfait_controlling_2016,eichler_electron_2017}, at a rate $\Gamma_\text{P} = 4g_0^2/\kappa$ for resonant spins. Because of the Purcell effect, spins have $T_1$ times in the order of a second or lower (see below) instead of hours~\cite{tyryshkin_electron_2012,bienfait_controlling_2016}, which makes signal averaging possible. Another specificity is the use of a superconducting parametric amplifier~\cite{macklin_nearquantum-limited_2015} to amplify the echo signal. This results in detection sensitivities in the $10-10^3~\text{spins}/\sqrt{\text{Hz}}$~\cite{bienfait_reaching_2016,probst_inductive-detection_2017,ranjan_electron_2020} range. On the other hand, the use of planar micro-resonators leads to large spatial inhomogeneity of the $B_1$ field and thus also of $g_0$, as shown in Fig.~\ref{fig:strainedbismuth}(f). The resulting spatially-inhomogeneous Rabi frequency and spin relaxation times lead to a complex echo response when subject to varying pulse amplitudes and repetition rates of the measurement~\cite{ranjan_pulsed_2020}. Some of these complex responses are described further below.

\begin{table}[t!]
\caption{Resonator parameters}
\label{table1}
\begin{center}
  \begin{tabular}{ | c | c | c | c | }
    \hline
     & Res1 & Res2 & Res3 \\ 
     \hline
    $w$ ($\mu$m) & 5 & 2 & 1 \\ \hline 
    $l$ ($\mu$m) & 700 & 450 & 450 \\ \hline  
    $\omega_0/2\pi$ (GHz) & 7.338 & 7.402 & 6.945\\ \hline
    $Z_0~(\Omega)$  & 40 & 40 & 45 \\ \hline
     $\kappa_\text{i}$ (s$^{-1}$) at $B_0=0$ & $4.6 \times 10^5$ &  $4.6 \times 10^5$ &  $ 5.5 \times 10^5$ \\ \hline
    $\kappa_\text{c}$ (s$^{-1}$)  & $4.6 \times 10^5$ & $4.6 \times 10^5$ & $1.5 \times 10^5$ \\ \hline
     $\kappa$ (s$^{-1}$) at $B_0=0$   & $9.2 \times 10^5$ & $9.2 \times 10^5$ & $7 \times 10^5$ \\ \hline   
  \end{tabular}
\end{center}
\end{table}

\section{Spectroscopy} 
\label{sec:sepctroscopy}

The spectrum of bismuth donors in silicon coupled to an aluminum superconducting micro-resonator patterned on top was measured recently~\cite{pla_strain-induced_2018} in a geometry very similar to the one used in our device. The spectrum was shown to be strongly distorted by strain in the substrate resulting from differential thermal contractions between the resonator and the silicon. The aim of this section is to support these earlier findings with complementary measurements performed on resonators having a different geometry.

\begin{figure*}[t!]
  \includegraphics[width=2\columnwidth]{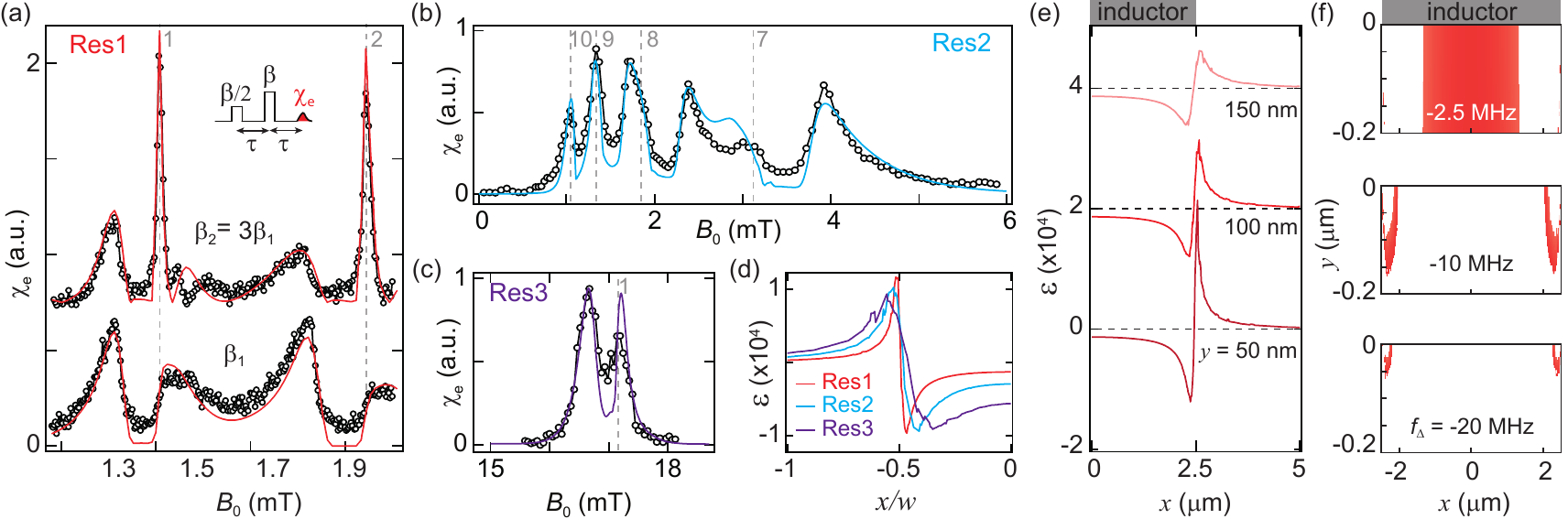}
  \caption{\label{fig:Spectrum}
  Strain spectroscopy. (a-c) Measured (symbols) and numerically modeled (solid lines) echo detected spin spectroscopy for three resonators. Hahn echo sequence consist of square pulses of amplitudes $\beta/2,~\beta$ which are each $2~\mu$s long and averaged at a repetition rate of 10~s (Res1), 2~s (Res2) and 0.5~s (Res3). Area under the echo is labeled as $\chi_e$. Vertical dashed lines mark the position of spin transitions in absence of any strain. The transition numbers ($i=1,~2,...,10$) are marked as described in Fig.~\ref{fig:strainedbismuth}(d). In panel (a), top (bottom) measurement is done at a pulse amplitude $\beta_2=15 \times 10^6 ~\text{s}^{-1/2}$ ($\beta_1 = 5 \times 10^6 ~\text{s}^{-1/2}$). (d) Comparison of hydrostatic component of strain $\varepsilon = (\epsilon_{xx} + \epsilon_{yy} +\epsilon_{zz})/3$  at a fixed depth $y=75$~nm across resonators and (e) at different depths in Res1. Curves have been offset by $2\times 10^{-4}$ for clarity. (f) Inferred location of spins (from COMSOL simulation) for different $f_\Delta$ within a square bandwidth of $\pm 0.2f_\Delta$ in Res1.   }
\end{figure*}

Strain arises during cooldown from room temperature, at which the device is fabricated, to milliKelvin temperatures, at which it is measured. The strain tensor $\varepsilon$ can be calculated using finite element simulations from COMSOL, using the known parameters for the differential thermal contraction of silicon relative to aluminum (see Appendix~\ref{app:strain}). From the strain tensor, the frequency shift $f_\Delta(\varepsilon)$ is computed using Eq.~\ref{eq:fDeltastrain}. An essential aspect of the present measurements is that because of the strain gradient, $f_\Delta(x,y)$ depends on the location $(x,y)$ of the donor. It also strongly depends on the width of the inductive wire, as can be seen in Fig.~\ref{fig:Spectrum}(d).

The computed strain-induced frequency shift $f_\Delta(x,y)$ is shown in Fig.~\ref{fig:strainedbismuth}(e) for the geometry of Res1. The values are in the $1 - 10\,\mathrm{MHz}$ range, much larger than both the resonator linewidth ($\kappa/2\pi \sim 150 \mathrm{kHz}$) and the $\sim 100~$kHz~\cite{weis_electrical_2012} linewidth due to the dipolar interaction with the bath of residual $^{29}$Si nuclear spins. Therefore, the bismuth donor spin spectrum is expected to be entirely dominated by strain shifts, as confirmed experimentally in the following. We also note that the strain and thus $f_\Delta(x,y)$ only weakly depends on $y$ unless $x$ is close to $\pm w/2$ [Fig.~\ref{fig:Spectrum}(e,f)].

Echo-detected spectra are measured with the sequence shown in the inset of Fig.~\ref{fig:Spectrum}(a) consisting of two square pulses of amplitude $\beta/2$ and $\beta$ and duration $t_p$, separated by a delay $\tau$, giving rise to an echo after a further delay $\tau$. The sequence is performed with a repetition time $\geq 3T_1$. The integral $\chi_e$ of the resulting echo is plotted as a function of $B_0$ in Fig.~\ref{fig:Spectrum} for the three resonator geometries. Moreover, for Res1, spectra are acquired with two different pulse amplitudes $\beta$.

We first notice that each spectral feature in Fig.~\ref{fig:Spectrum}(a) is composed of two split-peaks. Moreover, the right peak amplitude strongly depends on $\beta$, contrary to the left peak whose amplitude is barely changed. These observations can be qualitatively understood as follows. Spins located below the inductor have negatively shifted $f_\Delta$ and form the low-field peak. These have larger $g_0$ than those far away from the wire, and undergo Rabi angles close to $3\pi$ and $\pi$ for the two chosen $\beta$ values, leading to similar peak amplitudes. On the other hand, the high-field peaks are formed by weakly strained spins ($f_\Delta \sim 0$) located far from the wire and thus appear near the frequency of the unperturbed donor spectrum. Due to the weaker $g_0$, such distant spins undergo smaller Rabi angles than in the left peak, closer to $\pi$ and $\pi/3$ respectively, leading to very different echo amplitudes. The low-field (high-field) tails are formed by spins near the inner (outer) edges of the inductor that are also the most strained ones and closer to the sample surface.

Spectra acquired with Res2 and Res3 are presented in Fig.~\ref{fig:Spectrum}(b,c). We again observe a split-peak behavior, though there are significant overlaps in the case of Res2 due to the close spacing of the peaks. Another observation is that the peak splitting progressively increases from Res1 to Res3. For example, in Res3 the splitting is $\sim 0.5$~mT compared to $\sim 0.1$~mT in Res1. This is readily explained by the value of the hydrostatic strain at $x=0$, which is indeed $\sim 5$ times larger in Res3 than in Res1 [Fig.~\ref{fig:Spectrum}(d)]. Another visible feature in the measurements is the dependence of the linewidth on the transition. This is also consistent with strain broadening, which causes a frequency broadening $\Delta \omega = (\partial \omega/ \partial A) \Delta A $, uniform across all transitions (see Eq.~\ref{eq:lowfield}), and therefore a transition-dependent peak width of $ \Delta \omega / \geff(i)$ upon magnetic-field sweep. Note that Res2 frequency is larger than the ZFS, so that the split-peaks are inverted compared to Res1 and Res3, i.e. spins below the wire now give rise to the high-field peaks, and spins away from the wire to the low-field peaks.

To go beyond these qualitative arguments, a simple model is developed. It incorporates two physical effects: the spatial inhomogeneity of the strain shift $f_\Delta(x,y)$ and of the coupling constant $g_0(x,y)$, which leads to a spatially inhomogeneous Rabi angle $\theta(x,y) = 4 g_0 \beta /\sqrt{\kappa} t_p$ and Purcell relaxation rate $\Gamma_P (x,y)$. To compute the echo amplitude at a certain field $B_0$ due to transition $i$, we

- compute the corresponding shift $f_\Delta(i,B_0) = [\omega_0 - \omega_i(B_0)]/2\pi$

- Determine the spatial areas $\mathcal{A}(i,B_0)$ of donors whose strain shift is comprised

- Determine the total areas $\mathcal{A}(i,B_0)$ of the regions in the $xy$ plane, where the strain-induced shift is comprised between $f_\Delta(i,B_0) - \kappa/(2\pi)$ and $f_\Delta(i,B_0) + \kappa/(2\pi)$. Examples in Fig.~\ref{fig:Spectrum}(f) show that for larger values of $f_\Delta(i,B_0)$, the areas $\mathcal{A}(i,B_0)$ are closer to the edges of the wire and also to the sample surface. This is because strain increases close to the wire edges and substrate surface, as is apparent in Fig.~\ref{fig:Spectrum}(e,f); therefore donors that display a large strain-shift are only found in the area close to the wire edge and sample surface.

- Compute the echo amplitude due to transition $i$ as

\begin{multline}
    \label{eq:Ae}
    \chi_e(i,B_0) =2 \frac{\sqrt{\kappa_\mathrm{c}}}{\kappa} \iint_{\mathcal{A}(i,B_0)} dx dy l \rho(y) g_0(x,y) \\  \times [1-\exp(-\Gamma_\text{P}(x,y) t_\text{rep})] \sin^3[\theta(x,y)],
\end{multline}

\noindent where $\rho(y)$ is the implanted bismuth donor density and $l$ the wire length~\cite{ranjan_pulsed_2020}.

- We account for a possible inhomogeneity of the strain field by convolving the result with a Gaussian $\mathcal C(x)= \frac{1}{\sigma_B \sqrt{2\pi}} \exp(-x^2/2 \sigma_B^2)$ of  standard deviation $\sigma_B$:

\begin{equation}
    \tilde{\chi}_e (i,B_0)= \int \chi_e(i,B) \mathcal C (B_0 - B) dB.
\end{equation}

- The resulting echo amplitudes are finally summed over all transitions, with individual weights $p_i$ :

\begin{equation}
    \chi_e (B_0)= \sum_i p_i \tilde{\chi}_e (i,B_0).
\end{equation}

This model is slightly more complete than the one used in~\cite{pla_strain-induced_2018}, as it takes into account the spectrum distortions due to the finite repetition time and to the spatial distribution of the Purcell relaxation rate~\cite{ranjan_pulsed_2020}. The adjustable parameters are the local strain inhomogeneity $\sigma_B$ and the relative weights $p_i$ of the various transitions. Moreover, we allow ourselves to slightly adjust the Al deposition temperature in the strain calculation for each resonator geometry to match the measurements best. In ~\cite{pla_strain-induced_2018}, the implanted donor density $\rho(y)$ was moreover multiplied by the probability that a donor is ionized due to the Schottky barrier formed between the aluminum and the silicon, which leads to the formation of a depletion area below the wire. Here, we find a better agreement if we consider that the depletion area is of negligible extent, as shown in the Appendix~\ref{app:schottky}. This could be due to the existence of positive charges in the native silicon oxide layer, or to an exceedingly long time to reach charge equilibrium. Based on this observation, we neglect entirely the possible existence of a Schottky barrier in all the calculations described below. More work would be needed to determine the characteristics of the contact interface between aluminum and silicon in our device.

The result of the spectrum calculations is shown in Fig.~\ref{fig:Spectrum} for all resonator geometries, after optimization of the adjustable parameters. We find that a good agreement is obtained for all resonator geometries by taking $\sigma_B = 0.2 f_\Delta(i,B_0)/\geff $, suggesting that the strain tensor displays some amount of fluctuations around the value obtained by modelling the thin-film and substrate with a continuous medium theory. The next section will prove that this $\sim 20\%$ inhomogeneity of the strain field is actually local and happens on a scale of $\sim 50$~nm, which may be linked to the poly-crystalline nature of the aluminum thin-film. 

The model reproduces semi-quantitatively the two spectra measured on Res1 for different pulse amplitudes, for an initial temperature of $300$\,K. It also reproduces most of the spectral features measured for Res2 and Res3, for an deposition temperature of respectively $270$\,K and $250$\,K, reasonably close to room-temperature. The difference in deposition temperature may be due to the fact that the native silicon oxide is not included in the strain calculation. Interestingly, we need to assume different relative individual weights $p_i$ to reproduce the spectra: $p_1 = 1,~p_2 = 1.2$ for Res1, and $p_{10}= 1,~p_9 = 1.2,~ p_8 = 1.5,~ p_7 = 1.8$ for Res2, whereas all $p_i$ should be approximately equal at thermal equilibrium at 20\,mK. An out-of-equilibrium distribution of hyperfine states is not surprising, considering that the phonon relaxation rates in-between these levels become exceedingly long at low temperatures, whereas a variety of mechanisms (stray infrared radiation, carrier injection, ...) can cause some level of hyperpolarization in the hyperfine manifolds~\cite{sekiguchi_hyperfine_2010,albanese_radiative_2020}.

The quantitative understanding of the spectra achieved for various resonator geometry and pulse amplitudes confirms the results presented in \onlinecite{pla_strain-induced_2018}, stating that strain shifts are the dominant broadening mechanism in this kind of device. The refined analysis enables us to evidence two new features: 1) the strain tensor appears to fluctuate by $\sim 20\%$ around the value determined by finite-element modeling with a continuous-medium approach; it is tempting to link this inhomogeneity with the polycrystalline nature of the aluminum film, which is not taken into account in the modelling 2) the population distribution in the hyperfine manifold appears to be out-of-equilibrium. In the rest of this work, we leverage the correlation between position in the spectrum (corresponding to a given strain shift $f_\Delta(B_0)$ and spatial location [the area $\mathcal{A}(i,B_0)$] to obtain spatially-resolved information on the donor coherence properties and therefore on the noise they experience. We, however, note that the spatial resolution is mainly in the lateral dimension $x$, since strain is only weakly dependent on $y$ except close to the wire edges.

\section{Instantaneous diffusion}
\label{sec:ID}

In this section, we use the phenomenon of Instantaneous Diffusion (ID) to demonstrate that the $20\%$ strain inhomogeneity evidenced by the spectral measurements described in the previous section is local.

Various decoherence mechanisms for dilute paramagnetic impurities arise through dipolar interactions with other spins. Dipolar interactions between some `central spins’ and its neighbours can be refocused using a Hahn echo sequence $\pi/2 -\tau - \pi - \tau - echo$, however, if the neighbouring spins also fall within the bandwidth of the $\pi$ pulse, they are also flipped and the sign of the dipolar interaction term is not reversed. The effect of such dipolar interactions between similar spins can be probed by repeating the Hahn echo sequence using a smaller refocusing angle $\theta$, which can be interpreted as flipping only a fraction of spins. At the cost of a reduced echo intensity (due to the reduced probability $\sin^2(\theta/2)$ of flipping the central spin), the dipolar interactions with its neighbours is correspondingly refocused. The resulting coherence decay time constant, $T_2$, thus follows~\cite{schweiger_principles_2001} :

\begin{equation}
    \label{eq:ID}
1/T_2 = \Gamma_\text{res} + \rho \gamma_\text{eff}^2 \frac{\pi \hbar \mu_0}{9\sqrt{3}} \sin^2 \left (\frac{\theta}{2} \right ),
\end{equation}
where $\rho$ is the number of spins per unit volume, and $\Gamma_\text{res}$ the residual decoherence rate due to other spectral diffusion or decoherence mechanisms. Measuring the dependence of $T_2$ on $\theta$ thus provides a convenient method to determine the sample concentration $\rho$. Note that local inhomogeneous broadening tends to suppress ID, if this broadening is comparable to or larger than the pulse excitation bandwidth. Here we measure ID for various pulse excitation bandwidth and extract some information about local inhomogeneous broadening. 

\begin{figure}[t!]
  \includegraphics[width=\columnwidth]{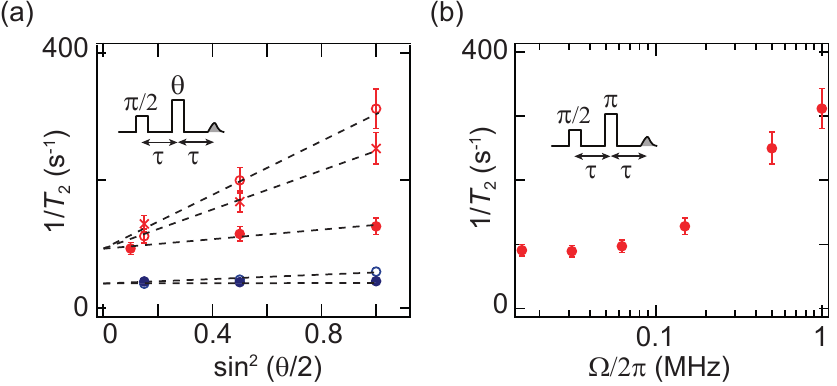}
  \caption{\label{fig:ID}
     Instantaneous diffusion in Res1. (a) Measured echo amplitude decay rate $T_2^{-1}$ as a function of $\sin^2 (\theta/2)$, $\theta$ being the Rabi angle achieved during the refocusing pulse, for the first transition (red symbols) and the fourth transition (blue symbols). Solid circles, crosses and open circles are data for refocusing pulse bandwidths of 150~kHz~($\kappa/2\pi$), 500~kHz and 1~MHz respectively. Dashed lines are linear fits to the data. (b) Decay rate $T_2^{-1}$ of the echo amplitude as a function of the pulse bandwidth $\Omega$, for transition $i=1$, and at a refocusing angle of $\pi$. Measurements are done with square pulses of duration $64~\mu$s, $32~\mu$s, $16~\mu$s, $2~\mu$s, and shaped pulses of duration $0.5~\mu$s and $1~\mu$s. }
\end{figure}

For that, we select spins located in the middle of the wire of Res1 (left peaks in Fig.~\ref{fig:Spectrum}(a), corresponding to $f_\Delta= -2.5~$MHz). Well-defined Rabi oscillations are obtained, enabling a quantitative knowledge of $\theta$ (see Appendix~\ref{app:Sx}). The coherence time $T_2$ is measured by varying the delay $\tau$ in a Hahn-echo sequence, and fitting an exponential $\exp(-2\tau/T_2)$ to the measured decay of the echo area $\chi_e$. The decoherence rate $1/T_2$ is shown in Fig.~\ref{fig:ID}(a) as a function of $\sin^2 (\theta/2)$, for different transitions and pulse excitation bandwidth $\Omega$. A linear dependence is found in all cases, indicative of ID. For a given bandwidth $\Omega/(2\pi) = 1$\,MHz, the ratio of slopes between the two transitions matches the expected $\gamma_\text{eff}^2$ dependence characteristic of ID (Eq.~\ref{eq:ID}). The slope is seen to strongly depend on $\Omega$ [Fig.~\ref{fig:ID}(a)], which we interpret as evidence for inhomogeneous broadening on a length scale corresponding to the distance between neighboring donors in the same level (thus, susceptible to be flipped by the same control pulse), which in our case is $\sim 50\,\mathrm{nm}$. Note that because of the limited microwave power, it was impossible to increase $\Omega/2\pi$ above $1$\,MHz, but ID keeps getting stronger up to that value, indicating that the broadening can be as high as $ \geq 1~$MHz.

Sources of local broadening include the Overhauser field due to the residual $^{29}\mathrm{Si}$ nuclear spins, which should be of order $100$\,kHz given the $5\cdot 10^{-4}$ relative $^{29}\mathrm{Si}$ concentration, significantly lower than what is needed to explain the data in Fig.~\ref{fig:ID}(b). Local electric field variations would need to be excessively large to explain our data and are thus unlikely. We believe that local variations of the strain tensor are the most likely explanation for our data, especially since strain inhomogeneity is also required to fit quantitatively the spectra as explained in the previous section. 

Using Eq.~\ref{eq:ID}, we extract from the data of Fig.~\ref{fig:ID}a (at $\Omega/2\pi = 1~$MHz) a concentration $\rho = 5 \times 10^{14}~\text{cm}^{-3}$ for bismuth donor spins in level $|4,-4\rangle$. This value is still an order of magnitude lower than the one expected from the nominal average bismuth concentration assuming a thermal distribution within the 9 hyperfine levels of the ground state manifold. We attribute this difference to a combination of finite donor activation yield~\cite{weis_electrical_2012}, insufficiently large $\Omega$ to fully compensate the strain inhomogeneity, and out-of-equilibrium distribution of the donors within the hyperfine manifold (as seen in Section IV).

\section{Decoherence due to magnetic noise}
\label{sec:magneticNoise}
We now leverage the transition- and magnetic field-dependent $\geff(i)$ [see Fig.~\ref{fig:strainedbismuth}(d) and Eq.~\ref{eq:lowfield}] as well as correlation between spatial position and strain-induced frequency shift to quantify the magnetic noise seen by bismuth donor spins, and get information about its spatial origin. In the following, all measurements are done with pulses limited by the resonator bandwidth ($\Omega/2\pi \sim 0.15~\mathrm{MHz}$), such that effect of ID is less than $20\%$ (see Fig.~\ref{fig:ID}). Moreover, we adjust the repetition time to be always longer than $3 T_1$ (see Appendix~\ref{app:Sx}), so that the spins are fully polarized for each measurement discussed below.

\begin{figure}[t!]
  \includegraphics[width=\columnwidth]{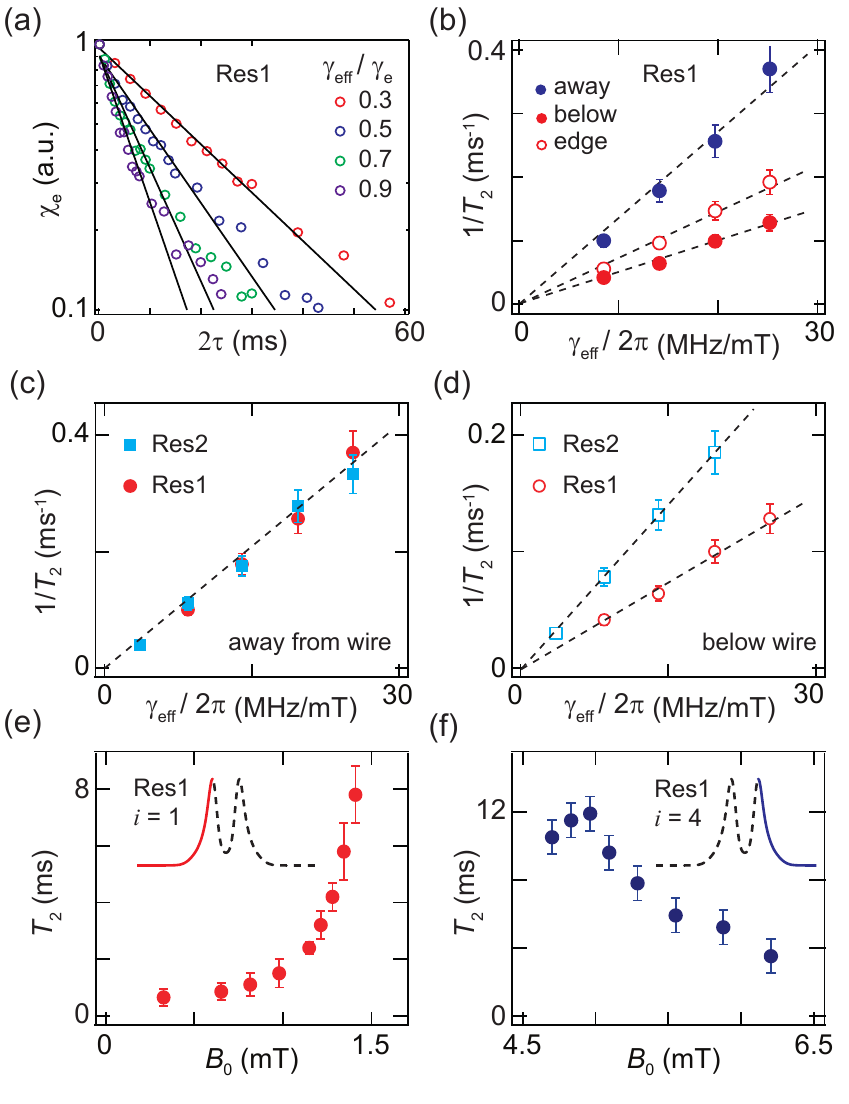}
  \caption{\label{fig:MagneticNoise}
   Spatial magnetic noise. (a) Echo integral $\chi_e$ as a function of the delay $2\tau$ between the $\pi/2$ pulse and the echo, measured with Res1 for transitions $i=1-4$, at fields $B_0$ corresponding to the same $f_\Delta = - 2.5$\,MHz (i.e., $\langle x \rangle \sim 0.5 \mu \mathrm{m})$. Open circles are the data, and solid lines are exponential fits yielding the decay rate $T_2^{-1}$. (b) Extracted echo amplitude decay rate $T_2^{-1}$ with Res1 as a function of $\gamma_\text{eff}$ for transitions $i=1-4$ and $f_\Delta = -2.5$\,MHz (spins below the wire, red dots), $f_\Delta = - 4$\,MHz (spins closer to the wire edge, red circles), and $f_\Delta = 0$\,MHz (spins far away from the wire, blue dots). The dashed lines are linear fits. (c-d) Comparison between resonators. Echo amplitude decay rate $T_2^{-1}$ as a function of $\gamma_\text{eff}$ measured with Res1 (transitions $i=1-4$, red dots) and with Res2 (transitions $i=6-10$, blue squares). Panel (c) shows the data for $f_\Delta = 0$ (spins far away from the wire), and panel (d) for spins below the wire. (e) Coherence time $T_2$ measured with Res1 as a function of $B_0$ on the low-field peak of the $i=1$ transition. (f) Coherence time $T_2$ measured with Res1 as a function of $B_0$ on the high-field peak of the $i=4$ transition.}
\end{figure}

Using resonator Res1, we  first compare the coherence time $T_2$ obtained across different transitions, from donor spins located in a region characterized by a given strain shift $f_\Delta$.  Figure~\ref{fig:MagneticNoise}(a) shows the Hahn-echo integrals $\chi_e(2\tau)$ on transitions $i$= 1 to 4, at $f_\Delta = - 2.5$\,MHz $(\langle x \rangle \sim 0.5~\mu$m). Exponential decays are systematically observed. The extracted exponential decay constant $T_2$ shows an increase from 7.5~ms ($i=1$, $\geff/\gamma_e = 0.9$) to 24~ms ($i=4$, $\geff/\gamma_e = 0.3$) [see Fig.~\ref{fig:MagneticNoise}(b)]. The same comparison is performed for two other values of $f_\Delta$, corresponding to spins below the wire but not in the middle ($f_\Delta=-4$~MHz, $\langle x \rangle \sim 1.5~\mu$m), and as far from the wire as possible ($f_\Delta \simeq 0,~ \langle x \rangle  \rightarrow \infty$). As seen in Fig.~\ref{fig:MagneticNoise}(b), we find that $T_2^{-1}$ linearly depends on $\geff$ for each of the three locations, with linear extrapolation at $\geff=0$ very close to $0$. This establishes that the dominant source of decoherence on the transitions considered here is of magnetic origin, for all three spin locations.

\begin{figure}[t!]
  \includegraphics[width=\columnwidth]{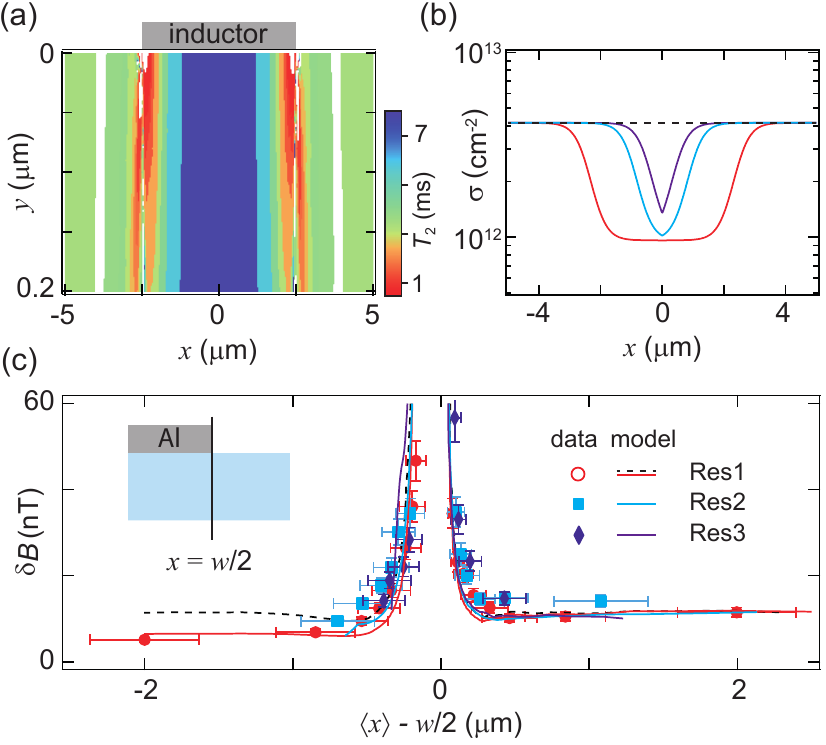}
  \caption{\label{fig:MagneticNoiseSpatialDependence}
   Spatially-dependent magnetic noise. (a) Measured coherence time $T_2$ of the first transition ($i=1$) versus the inferred spin locations in Res1. Obtained from $T_2 (B_0)$ and $\mathcal{A}(B_0)$, the seemingly limited $y$ resolution in $T_2 (x,y)$ is due to quasi 1-D strain shifts (see text for more details. (b) Reconstructed surface magnetic impurity distribution $\sigma$ as a function of the position $x$ with respect to the wire center for the three resonators. (c) Magnetic noise magnitude for the three different resonators as a function of the average spin position with respect to the wire edge. Measurements are colored symbols, and the model described in the text is shown by solid lines. The dashed black curve is computed for Res1 assuming a uniform density $\sigma = 4 \times 10^{12}~\text{cm}^{-2}$.  }
\end{figure}

Without having to assume any decoherence model, we can re-express the measured coherence times $T_2$ into an effective magnetic noise by computing $\delta B = 2 \pi / (\geff T_2)$. As seen from line slopes in Fig.~\ref{fig:MagneticNoise}(b), we observe a marked dependence of $\delta B$ on the location of the probed donor spins relative to the wire and substrate surface. Complementary measurements are provided by comparing $\delta B$ in Res1 and Res2. As seen in Fig.~\ref{fig:MagneticNoise}(c), $\delta B$ is found to be the same in Res1 and Res2 for spins away from the wire ($x \rightarrow \infty$, $f_\Delta \simeq 0$), whereas $\delta B$ is twice larger in Res2 than in Res1 for spins at $\langle x \rangle \sim 0.5~\mu$m, right below the wire. To obtain a comprehensive measurement of $\delta B$, we measure $T_2$ as a function of $B_0$ on the edge of the low-field peak on transition $i=1$ [Fig.~\ref{fig:MagneticNoise}(e)] and on the edge of the high-field peak of transition $i=4$ [Fig.~\ref{fig:MagneticNoise}(f)], as the latter has negligible overlap with other transitions. $T_2$ is seen to vary by as much as one order of magnitude, with the magnetic noise $\delta B$ increasing in strength for spins closer to the wire edge and to the substrate interface. Similar results are obtained with Res2 and Res3. With the knowledge of $\mathcal{A}(i,B_0)$, we deduce a spatial map of the decoherence, which is shown in Fig.~\ref{fig:MagneticNoiseSpatialDependence}(a, c). 

The observation that the magnetic noise $\delta B$ causing bismuth donor spin decoherence is spatially dependent is one of the main results of our work, as it has important implications on the possible physical origin of this noise. Spectral diffusion caused by a spin bath uniformly distributed throughout the bulk of the sample (such as residual $^{29}\mathrm{Si}$ nuclear spins, or any other paramagnetic impurities including the donors themselves) would act uniformly on all donor spins and is therefore incompatible with our data. Another usual decoherence mechanism is noise in the magnet applying $B_0$~\cite{muhonen_storing_2014}; it can also be ruled out in our case, for the same reason. 

On the other hand, the spatial dependence of $\delta B$ and its increase close to the sample surface is a strong indication that decoherence is dominantly caused by the interaction of the donor spins with a spin bath residing at the silicon/silicon oxide interface, as proposed in Ref.~\onlinecite{de_sousa_dangling-bond_2007} and reported in Ref.~\onlinecite{schenkel_electrical_2006,paik_t_1_2010}. Such a spin bath could consist for instance of dangling bonds (the $P_\mathrm{b}$ defects). 

Decoherence of a spin located at a depth $d$ below a noisy surface containing a surface density $\sigma$ of fluctuating spins was considered in several contexts. One model~\cite{de_sousa_dangling-bond_2007} considers the decoherence of donors in silicon caused by spin-lattice relaxation of paramagnetic $P_\text{b}$ defects at the silicon/silicon oxide interface. It predicts that the echo of an isolated donor with gyromagnetic ratio $\gamma$ should have a Gaussian decay $\chi_e(2\tau)/\chi_e(0) = \exp [-(2\tau / T_2)^2]$, with $T_2 = K / [\gamma \delta b(d)] $, $\delta b (d)$ being the standard deviation of the magnetic field generated by a distribution of random surface spins at the depth $d$, and $K \sim 10$ a dimensionless parameter linked to the specific modeling of the paramagnetic defects distribution. It can be shown that $\delta b (d)$ scales like $\sqrt{\sigma}/d^2$. Another (less specific) model assumes that the surface spins cause a magnetic noise with power spectrum $S(\omega) = S_0(d) /\omega^{\alpha}$. In our case, we have evidence that $\alpha \simeq 1$ (see Section~\ref{sec:DDandTemp}), which should also lead to a Gaussian Hahn-echo decay with $T_2 = 2\pi/[\gamma \sqrt{S_0(d)}]$~\cite{ithier_decoherence_2005,yoneda_quantum-dot_2018}. Because of the $1/f$ divergence at low frequencies, in this model it is not strictly speaking possible to define a magnetic field noise standard deviation; nevertheless, after introducing proper infra-red cutoff frequencies for the noise, one finds that within a factor of order unity, $S_0(d) \sim \delta b^2(d) $. In summary, both models predict a Gaussian-shaped decay, with $T_2 \sim 2\pi / [\gamma \delta b(d)] $ within a factor of order unity which we will disregard in the following. 

In our device, the interface cannot be considered homogeneous because of the metallic wire, and two areas can be distinguished. Away from the wire, the interface is between the silicon substrate and the native silicon oxide; whereas below the wire the interface is more complex with the $50$\,nm-thick aluminum wire on top of the Si/SiO2 interface. This may lead to differences in surface spin density, which we take into account in a minimal model where the surface spin density evolves from $\sigma_1$ away from the wire to $\sigma_2$ below the wire, with a smooth interpolation of characteristic width $500$\,nm. We numerically estimate the resulting magnetic noise, by discretizing the surface spins into point dipoles of magnitude Bohr magneton ($\mu_\text{B}$) with random orientation, and summing the squares of the magnetic field component along the applied field $z$ direction, thus obtaining the rms magnetic noise $ \delta b(x,y) $. The echo amplitudes for a given $f_\Delta$ and delay $2\tau$ are then numerically calculated using

\begin{multline}
    \label{eq:T2}
    \chi_e(f_\Delta,\tau) =2 \frac{\sqrt{\kappa_\mathrm{c}}}{\kappa} \iint_{\mathcal{A}(i,B_0)} dx dy l \rho(y)  g_0(x,y) \\  \times [1-\exp(-\Gamma_\text{P}(x,y) t_\text{rep})] \exp[-(2\tau/T_2(x,y))^2],
\end{multline}

\noindent with $1/T_2(x,y) = \Gamma_\text{non} + \delta B(x,y) \geff/2\pi$, the non-magnetic coherence time $1/\Gamma_\text{non} \sim 0.2~$s being estimated from measurements around the CT (see Section~\ref{sec:non-magneticnoise}). Despite the decay of each pixel being Gaussian, we find that the resulting sum decays closer to an exponential, in agreement with the measurements [Fig.~\ref{fig:MagneticNoise}(a)]. We then fit this exponential and extract the model-predicted $T_2(f_\Delta)$. We note that according to this model, the measured $T_2(f_\Delta)$ contains contribution of vastly-different depth-dependent spin-echo decay curves, since strain shift $f_\Delta$ provides limited resolution in $y$ [Fig.~\ref{fig:MagneticNoiseSpatialDependence}(a)]. 

In order to compare the model to the data, the model-predicted $T_2(f_\Delta)$ can be converted into an effective magnetic noise $ \delta B(f_\Delta)= 2 \pi/[ \geff T_2(f_\Delta)]$. Figure ~\ref{fig:MagneticNoiseSpatialDependence}(c) shows the experimental and theoretical $\delta B$ as a function of the average lateral position of the donors $\langle x \rangle$ relative to the wire edge, for the three resonators. A semi-quantitative agreement for all resonators is obtained for $\sigma_1 = 4 \cdot 10^{12}~\mathrm{cm}^{-2}$ and $\sigma_2 = 10^{12}~\mathrm{cm}^{-2}$ [see Fig.~\ref{fig:MagneticNoiseSpatialDependence}(b)]. The numerical calculations in particular capture the diverging behavior of $T_2^{-1}$ near the wire edge as strained donors, on average, are closest to both the surface and wire edges. For Res1, the dashed curve shows the result of the calculation assuming a constant density $\sigma_1 = \sigma_2$; we see that the longer $T_2$ time measured below the $5~\mu \mathrm{m}$ wire cannot be reproduced in this way. Defect densities of $\sim 10^{12}\,\mathrm{cm}^{-2} $ are consistent with typical defect density at the Si / SiO2 interface~\cite{van_gorp_dipolar_1992}, and with the numbers reported in Ref.~\cite{de_sousa_dangling-bond_2007}. These numbers are also comparable to the ones inferred from flux-noise measurements in superconducting circuits (between $1 - 5 \times~10^{13}\, \mathrm{cm}^{-2}$)~\cite{koch_model_2007,sendelbach_magnetism_2008,braumuller_characterizing_2020}. 

In all the above discussion, we disregarded the magnetic response of the superconducting aluminum wire, which may have an impact on the field $\delta B$ generated by the layer of surface dipoles due to the Meissner screening of the magnetic field. This assumption was confirmed by calculations described in the Appendix~\ref{app:Meissner}, which show that the impact of the Meissner screening is negligible in our geometry and cannot explain the spatial dependence of $T_2$. 

It is interesting to compare our measurements (Fig.~\ref{fig:MagneticNoise}) to previous decoherence studies of near-surface spins in isotopically purified silicon. Antimony ions of concentration $10^{16}~\text{cm}^{-3}$ implanted at an average depth of $ 100~$nm in samples without metal electrodes, and measured using bulk EPR spectrometer showed $T_2 = 1~$ms~\cite{schenkel_electrical_2006}. This value is quite similar to the $3$\,ms measured far from the wire, supporting a similar decoherence mechanism in both cases. In a different experiment, individual phosphorus donors~\cite{muhonen_storing_2014} were implanted at much lower depth ($15-20$\,nm), and measured in devices with close spacing of metal electrodes. Using the $1/d^2$ scaling of the noise standard deviation, we conclude that such a phosphorus donor in our device would have a coherence time around $100~\mu \mathrm{s}$ at most, whereas $T_2 = 1~\mathrm{ms}$ was reported~\cite{muhonen_storing_2014}. This difference can be attributed to a better interface quality, and/or to the larger value of $B_0$ (exceeding one tesla), which leads to a complete polarization of impurity spins at $10$\,mK, contrary to our measurements done at low fields. Deconvolving the two effects would be interesting and could be the subject of future work.

The observation that the surface spin density is reduced below the aluminum wire compared to its value away from the wire may be an indication that aluminum has a passivating effect on dangling bonds at the silicon/silicon oxide interface. In this context, it is interesting to note that aluminum deposition has been shown to affect the underlying silicon oxide layer, leading to a reduction of its thickness at the expense of the growth of an aluminum oxide layer~\cite{lim_electrostatically_2009,spruijtenburg_fabrication_2018}. More work would be needed to confirm this observation and understand its physico-chemical origin.

\section{Decoherence due to non-magnetic noise} 
\label{sec:non-magneticnoise}

\begin{figure}[t!]
  \includegraphics[width=\columnwidth]{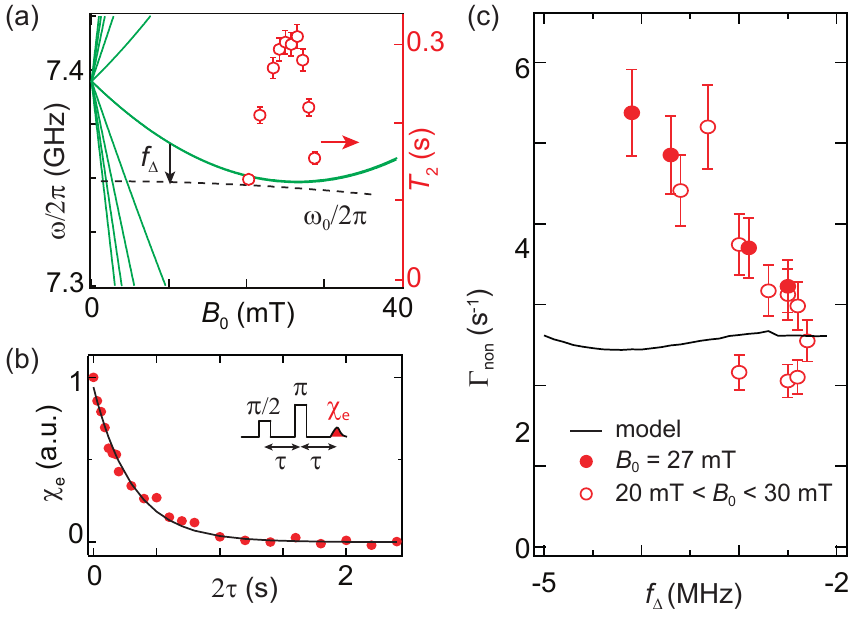}
  \caption{\label{fig:ElectricNoise}
  Non-magnetic noise characterization at CT in Res1. (a) Calculated spin transition frequencies (green lines), measured resonator frequency (dashed line, left axis), and measured coherence time $T_2$ (symbols, right axis) as a function of $B_0$. (b) Echo integral $\chi_e$ as a function of the delay $2\tau$ between the $\pi/2$ pulse and the echo. Red dots are data, and solid line is an exponential fit yielding a coherence time $T_2 = 300$\,ms. Magnitude detection is used to circumvent phase noise from the measurement setup. (c) Echo decay rate $\Gamma_\text{non}$ as a function of $f_\Delta$. Red dots are measurements at $B_0 = 27~$mT performed over several runs, red circles are extracted from the data in panel (a). Black solid line is the result of the model described in the main text.}
\end{figure}

We study the sensitivity of spins to non-magnetic noise sources by suppressing the dominant magnetic noise contribution near the CT in Res1. As expected from Fig.~\ref{fig:MagneticNoise}(c), $T_2$ increases with decreasing $\geff$, and reaches a maximum value of $0.3~$s when $\geff=0$~[Fig.~\ref{fig:ElectricNoise}(a)]. To examine if $T_2$ is influenced by second order magnetic fluctuations, we estimate its upper bound using $1/T_2 = \Gamma_\text{non} + (k_1/2\pi) (d\omega/d B_0) + ( k_1/2\pi)^2 (d^2 \omega/d^2 B_0)$, where $\Gamma_\text{non}$ describes the decoherence from non-magnetic origin and $k_1$ is a position-dependent average magnetic noise fluctuation. With $k_1 \sim 5~$nT for spins below the inductor~[Fig.~\ref{fig:MagneticNoise}(b)], we obtain $ ( k_1/2\pi)^2 (d^2 \omega/d^2 B_0) = 2.5 \times 10^{-6}~\text{s}^{-1}$, many orders of magnitude lower than the measured $1/T_2 = 3~\text{s}^{-1}$. Therefore, the magnetic noise evidenced in Section \ref{sec:magneticNoise} is not causing the decoherence measured at the CT.

Even though $\geff = 0$ at the CT, flip-flops with neighboring spins are still possible since the transverse spin matrix element $\langle 0 | S_x | 1 \rangle = 0.25 $ is non-zero. This decoherence mechanism (called direct flip-flop~\cite{tyryshkin_electron_2012}) is thus not suppressed at the CT. It was found to be dominant in measurements on bulk-doped bismuth donors in silicon ~\cite{wolfowicz_atomic_2013}, with an observed scaling $T_2 \simeq 10^3 \mathrm{s}\cdot \mu \mathrm{m}^{-3} /\rho $, where $\rho$ is the implanted donor concentration expressed in $\mu \mathrm{m}^{-3}$. In our device, this process would limit $T_2$ to $\sim 25~\mathrm{ms}$ for our $\rho \sim 4 \times 10^4 \mu \mathrm{m}^{-3}$; it appears therefore to be strongly suppressed, likely due to the locally inhomogeneous strain shifts already evidenced in Section \ref{sec:ID}. Moreover, dynamical decoupling measurements at the CT (see Section~\ref{sec:DDandTemp}) are found to extend the coherence time, which is incompatible with a direct flip-flop mechanism. Overall, we thus conclude that decoherence by direct flip-flop is negligible in our measurements, even at the CT. It is also interesting to note that strain effectively protects the donors from decoherence, as it was also shown to increase the spin-lattice relaxation time~\cite{tenberg_electron_2019}. 

Because the CT happens at a fixed value of $B_0$, it is not possible to study the dependence of coherence time on $f_\Delta$ as thoroughly as was done for magnetic noise (see Section \ref{sec:magneticNoise}). Nevertheless, partial data was obtained, in two different ways. First, due to field-sweep induced vortices in the superconducting aluminum and variations from one experimental run to another, the resonator frequency $\omega_0$ at the CT ($B_0 = 27~$mT) was found to vary over time and from run to run, enabling to measure $T_2$ for various values of $f_\Delta$ [full circles in Fig.~\ref{fig:ElectricNoise}(c)]. Another dataset is obtained independently, in the course of a single run, using measurements of $T_2$ as a function of $B_0$ around the CT [see Fig.~\ref{fig:ElectricNoise}(a)]. Because of the known dependence of both the bismuth unperturbed transition frequency [green solid line in Fig.~\ref{fig:ElectricNoise}(a)] and the resonator frequency (dashed black curve in Fig.~\ref{fig:ElectricNoise}a) on $B_0$, these measurements probe different values of $f_\Delta$. Moreover, we can estimate the magnetic noise contribution, using the known transition slopes $\gamma_\text{eff}(B_0)$ and $k_1 \sim 5~$nT. Subtracting this contribution from the inverse measured coherence time, we obtain $\Gamma_\text{non} (f_\Delta)$ [open circles in Fig.~\ref{fig:ElectricNoise}(c)]. Both datasets are consistent with one another and show that $T_2$ at the CT decreases for donors located under the wire in more strained areas (ie, closer to the wire edges). These observations indicate a spatial dependence of the non-magnetic noise responsible for decoherence at the CT.

A likely phenomenon to explain our data is charge noise originating from the interfaces between silicon, silicon oxide, and aluminum oxide, which can produce Stark shifts of the hyperfine constant. $1/f$ charge noise is ubiquitous in silicon nanoelectronic devices. Its study was revived recently at low temperatures in the context of research on quantum dot spin qubits~\cite{petit_spin_2018,connors_low-frequency_2019}. 

As explained in Section~\ref{sec:Bidopants}, charge noise couples to the Bi donor spin because of the quadratic Stark effect $\Delta A(\mathbf{E})/A_0 = \eta E^2$ with $\eta = (-0.26 \pm 0.05) \times 10^{-3}~\mu \mathrm{m}^2 / \mathrm{V}^2$. In our device, the donors are subject to a residual electric field of order $E_\mathrm{r} \sim 0.1~$mV/nm due to the bulk boron doping. We can therefore estimate the coherence time caused by electric field fluctuations due to charge noise around the residual field $E_\mathrm{r}$ as $T_2 = 2 \pi / (10 \eta A_0 E_\mathrm{r} \sqrt{S_E}) $, assuming fluctuations with a $1/f$ spectrum $S_E(\omega) = S_E / \omega$, where $S_E$ is the electric field noise power spectral density at $1$\,Hz. Using electrostatic simulations, we can moreover express the electric field fluctuations $\sqrt{S_E}$ as if they were entirely caused by voltage fluctuations $\sqrt{S_V}$ of the aluminum gate electrode. We find that for our device, $\sqrt{S_V} = 1.5~$mV yields a coherence time similar to our measurements [see solid line in Fig.~\ref{fig:ElectricNoise}(c)]. This level of noise is significantly larger than the gate voltage noise measured in quantum dot devices $\sqrt{S_V} \sim 0.01$~mV ~\cite{petit_spin_2018,connors_low-frequency_2019}, which may indicate that $E_\mathrm{r}$ is larger than estimated, or that our electrostatic modelling is too crude. We stress that our simplistic reasoning is here only meant to show that typical charge noise levels may indeed limit the coherence time to the values that we measure at the clock transition.

\section{Additional noise characterization}
\label{sec:DDandTemp}

In this section, we present complementary measurements to further characterize the noise responsible for spin decoherence. 

First, we use dynamical decoupling to obtain further information about the noise power spectrum $S(\omega) $. Indeed, assuming that $S(\omega) $ scales as $\omega^{- \alpha}$, it can be shown~\cite{alvarez_measuring_2011,yuge_measurement_2011,medford_scaling_2012} that the coherence time measured in a dynamical decoupling sequence containing $N$ refocusing pulses should scale as $N^\gamma$,  with $\gamma = \alpha / (1+\alpha)$; measuring $\gamma$ therefore gives access to $\alpha$. 

\begin{figure}[t!]
  \includegraphics[width=\columnwidth]{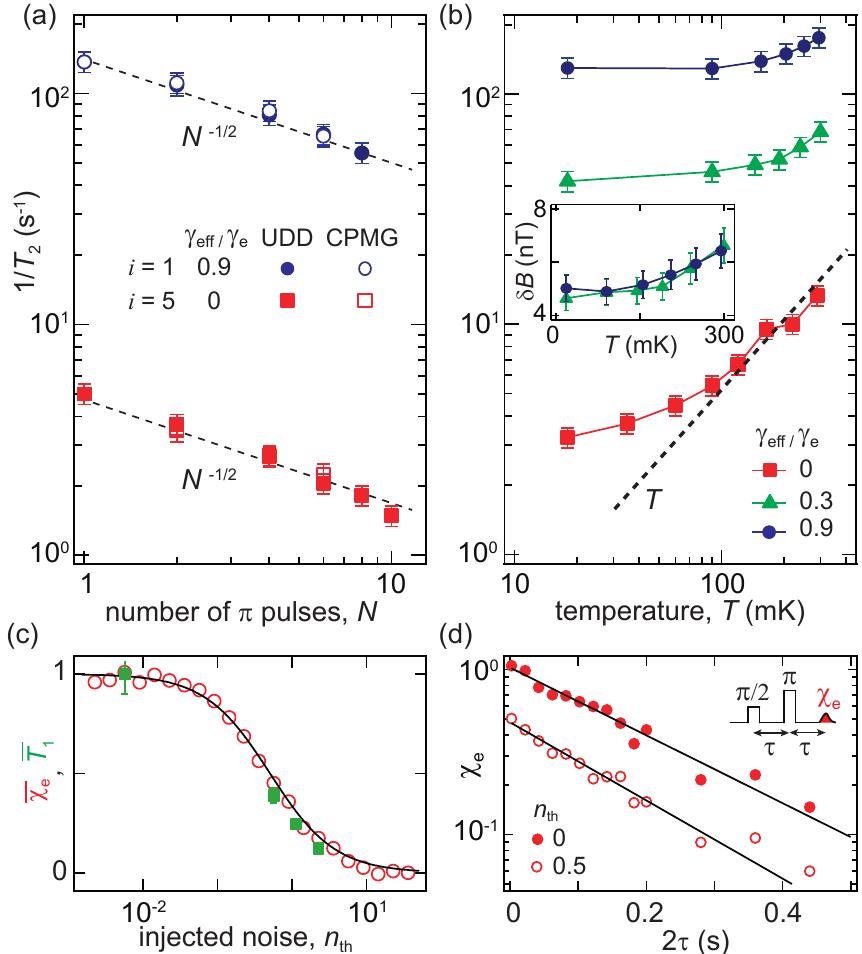}
  \caption{\label{fig:DD}
    Additional noise characterizations in Res1 at $f_\Delta = -2.5$~MHz (spins below the wire). (a) Echo decay rate $1/T_2$ as a function of the number of pulses in the dynamical decoupling sequence, on transition $i=1$ (blue) and $i=5$ (CT, red). Full (open) symbols correspond to the UDD (CPMG) sequence. Dashed line indicates the $N^{-1/2}$ scaling. (b) Echo decay rate $1/T_2$ as a function of sample temperature $T$, for transitions $i=1$ (blue), $i=4$ (green), and $i=5$ (CT, red). The dashed line is a guide to the eyes suggesting a $T^{-1}$ scaling at high temperatures of the decay times at the CT. Inset: $\delta B = 2\pi/(T_2\gamma_\text{eff})$ obtained by subtracting the contribution from non-magnetic noise (values measured at CT), as explained in the main text. (c) Normalized echo amplitude (circles) and spin relaxation time $T_1$ (squares) in the presence of injected thermal noise $n_\text{th}$. Symbols and line are, respectively. measurements and theory from Eq.~\ref{eq:purcellNoise}. (d) Echo integral $\chi_e$ as a function of the delay $2\tau$ at the $i=5$ transition (CT), without (red dots) and with (red circles) injection of $n_\text{th} = 0.5$ photons of thermal noise. Solid lines are exponential fits to the data, and show that the decay rate is not affected by thermal noise. }
\end{figure}

We perform coherence time measurements with the Uhrig Dynamical Decoupling (UDD), and Carr-Purcell-Meiboom-Gill (CPMG). The data are taken with Res1, at $B_0 = 1.4$\,mT (1st transition, giving access to the magnetic component of the noise), and $B_0 = 27$\,mT (CT, giving access to the charge noise). To obtain reliable data, it was necessary to remove the contribution of spurious echoes due to pulse imperfections, using phase cycling ~\cite{schweiger_principles_2001} (see Appendix~\ref{app:DD}). With both sequences, we find that $T_2$ increases with the number $(N)$ of $\pi$ pulses [Fig.~\ref{fig:DD}(a)], with an approximate $N^{1/2}$ scaling, both at the 1st transition and at the CT. This suggests that the noise power exponent $\alpha$ is close to $1$, indicating an approximate $1/f$ noise spectrum both for the magnetic noise and the charge noise. We note that a similar $T_2 \propto N^{1/2}$ scaling was observed in NV centers close to the diamond surface (therefore, probing surface magnetic noise)~\cite{myers_probing_2014}, and in semiconducting quantum dots subjected to charge noise originating from the silicon/silicon oxide interface~\cite{yoneda_quantum-dot_2018}. Magnetic noise originating from surface fluctuators is also commonly believed to be responsible for $1/f$ flux-noise in superconducting qubits~\cite{yoshihara_decoherence_2006,bylander_noise_2011}. As discussed earlier, the exponential shape of the echo decay observed on both transitions appears to be the result of an average of Gaussian decays with an inhomogeneous distribution of $T_2$ times due to the varying donor depth. 

We also measured the coherence time as a function of temperature $T$, at the CT ($T_\text{2,CT}$), the 1st ($T_\text{2,1st}$) and 4th ($T_\text{2,4th}$) transitions in Res1. As seen in Fig.~\ref{fig:DD}(b), $T_\text{2,CT}(T)$ shows a strong temperature dependence down to the lowest temperatures, whereas $T_\text{2,1st}(T)$ and $T_\text{2,4th}$ have a much weaker temperature dependence. We can estimate the temperature dependence of the purely magnetic noise by computing $T_{2,m}^{-1}(T) = T_\text{2,1st/4th}^{-1}(T) - T_\text{2,CT}^{-1}(T)$. Note that this subtraction has no rigorous justification since all measured $T_2$s are in fact averages over various Gaussian decays, for which the magnetic and charge decay rates do not add up as simply; nevertheless, $T_{2,m}^{-1}(T)$ has the merit to be directly accessible from the measurements. We see that $T_{2,m}^{-1}(T)$ is unchanged up to $200$\,mK, and slightly increases up to $300$\,mK. In contrast, $T_\text{2,CT}^{-1}(T)$ is 5 times stronger at $300$\,mK than at $20$\,mK. This different temperature dependence confirms the different physical origin of the noise at the 1st transition and at the CT. We also note that a strong temperature dependence in the 10-200mK range was observed for charge noise~\cite{connors_low-frequency_2019}, which further supports our interpretation. 

We also quantify the dependence of $T_2$ on the presence of thermal photons in the resonator, which are a common source of decoherence for superconducting qubits~\cite{bertet_dephasing_2005}. In that goal, we inject microwave noise in the input line by connecting the output of a pair of amplifiers in series at room temperature, while keeping the sample temperature at $20$\,mK. The average number of thermal noise photons $n_\text{th}$ is calibrated using the dependence of both the echo amplitude $\chi_e$ and energy relaxation time $T_1$ on $n_\text{th}$ in the Purcell regime~\cite{albanese_radiative_2020}

\begin{equation}
\label{eq:purcellNoise}
\frac{\chi_e(n_\text{th})}{\chi_e(0)}  = \frac{T_1(n_\text{th})}{T_1(0)}  =  \frac{1}{(2 n_\text{th}+1)}.
\end{equation}

The data showing $\chi_e$ and $T_1$ versus $n_\text{th}$ are shown in~[Fig.~\ref{fig:ElectricNoise}(c)], closely matching the theoretical prediction and enabling an absolute calibration of $n_\text{th}$. We then compare echo decays for $n_\text{th} = 0$ and $0.5$. For the latter, the thermal noise corresponds to an effective photon temperature of $0.35$~K and both echo amplitude and $T_1$ are halved. We extract similar $T_2$ values~[Fig.~\ref{fig:DD}(d)] and thus conclude that thermal radiation does not limit $T_2$ in our device.

\section{Conclusion}  \label{sec:conclusion}

We have performed spectral and coherence time studies of shallow-implanted bismuth electron spins in $^{28}\mathrm{Si}$-enriched silicon, using superconducting micro-resonators with different device geometries, and at temperatures between 20~mK and 300~mK. We quantitatively describe the spin spectra, which we show to be dominated by strain broadening specific to the sample geometry. Our measurements reveal the existence of locally inhomogeneous strain gradients, which largely suppress both instantaneous diffusion and direct flip-flop decoherence. 

The spatially-dependent frequency shifts caused by strain enable a spatially-resolved measurement of donor spin decoherence. Together with the transition-dependence of the effective gyromagnetic ratio, this makes it possible to pinpoint spectral diffusion due to surface paramagnetic impurities as the dominant decoherence mechanism. Our data enable some amount of spatial resolution of the surface spin density, which we estimate to be $10^{12} \mathrm{cm}^{-2}$ below the aluminum wire, and $4 \times 10^{12} \mathrm{cm}^{-2}$ away from the wire. These figures are comparable to typical dangling bond densities at the silicon/silicon oxide interface. The physico-chemical origin of the reduced defect spin density below the aluminum is unclear. This result, which deserves to be confirmed by further work, is also relevant for the understanding and mitigation of flux-noise in superconducting quantum circuits~\cite{braumuller_characterizing_2020}.

At the Clock Transition, the donors are no longer sensitive to magnetic noise. Our measurements suggest that the residual decoherence is caused by charge noise at the sample surface, still enabling to reach long $T_2$ values of up to $300$\,ms, which confirms that donors in silicon can be used for quantum devices despite the detrimental effect of interfaces~\cite{ranjan_multimode_2020-1}. 

To extend this study further, one possible direction is to increase the level of control over the donor charge state and strain shifts, using gate electrodes or piezoelectric materials. It would also be interesting to study the impact of fully polarizing the surface paramagnetic defects on the donor coherence time, by repeating similar measurements at higher magnetic fields $B_0$ than was done in this study - given the temperature of $10$\,mK, fields above $\sim 100$\,mT should be sufficient. More generally, the EPR spectroscopy of small ensembles of donor spins with spatial resolution appears as a powerful tool towards the development of spin-based quantum devices. 

\appendix
\section{Transition matrix elements}
\label{app:Sx}
\begin{figure}[t!]
  \includegraphics[width=\columnwidth]{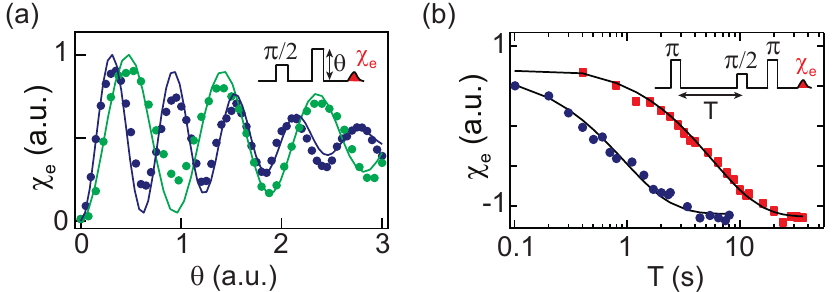}
  \caption{\label{fig:T1andRabi}
    (a) Measured (symbols) and numerically simulated (lines) Rabi oscillations at transitions $i=1$ (blue) and $i=4$ (green) with Res1. (b) Measured spin energy relaxation with inversion recovery sequence for the transitions $i=1$ (circles) and CT (squares). Lines are best fits yielding $T_1$ of $ 1 \pm 0.2~$s and $5.5 \pm 1~$s, respectively. The resonator loss rate $\kappa$ at CT is 30\% larger than the first transition due to magnetic field induced vortices. }
\end{figure}

Strong electronic-nuclear hyperfine interaction in Si:Bi leads to 18 $S_x$ transitions at low magnetic fields. In Fig.~\ref{fig:strainedbismuth}(b), we have labeled those as 1 and 10 for two non-degenerate, and 2 to 9 for eight quasi doubly-degenerate transitions. The latter can be recognized in the coupled basis as $|4,m-1\rangle \leftrightarrow |5,m\rangle$ and $|4,m \rangle \leftrightarrow |5,m-1\rangle$, $|m| \leq 4$. Interestingly two quasi-degenerate transitions are complimentary, i.e. sum of their $S_x$ matrix elements $\approx 0.5$. For example, the clock transition near $27~$mT has $S_x$ matrix element of $\langle 4,0| S_x |5,-1\rangle$ = $\langle 4,-1| S_x |5,0 \rangle$ = 0.25. In contrast, for non-degenerate transitions $|4,-4 \rangle \leftrightarrow |5,-5\rangle$ and $|4,4\rangle \leftrightarrow |5,5\rangle$, labeled 1 and 10 respectively in Fig.~\ref{fig:strainedbismuth}(b), it is $0.48$. Because matrix elements directly affect spin-photon coupling strength $g_0$, Rabi flops are transition dependent~[Fig.~\ref{fig:T1andRabi}(a)]. The same happens for spin energy relaxation times as our spins are in the strong Purcell limit~\cite{bienfait_controlling_2016}. $T_1$ measured using inversion recovery sequence for the first transition and CT are shown in Fig.~\ref{fig:T1andRabi} and found in excellent agreement with expected Purcell relaxation time $T_1 = \kappa/4 g_0^2$.

The echo amplitude response across different transitions are complex because of $g_0$ variations, which also affects the Rabi angle for a given pulse amplitude and polarization for a given repetition time of the experiment (see Eq.~\ref{eq:Ae}). For example, Rabi flops performed at $t_\text{rep}=4~$s on the fourth transition [Fig.~\ref{fig:T1andRabi}(a)] receives dominant echo contribution from $|4,-1\rangle \leftrightarrow |5,-2\rangle$ with higher matrix element of 0.32 (extrapolated $T_1 = 2.2$~s) and less so from $|4,-2\rangle \leftrightarrow |5,-1\rangle$ with matrix element of 0.18 (extrapolated $T_1 = 7~$~s). Such transition dependent changes in $g_0$ are accounted for in the numerical simulations.

\section{Strain spectroscopy simulation}
\label{app:strain}
We have performed finite element simulations of strain using COMSOL software. To this end, we input the temperature dependent thermal expansion coefficients of the materials, and their stiffness tensor coefficients~\cite{pla_strain-induced_2018}. Other input parameters are the thickness of Al ($50 \pm 5$~nm measured by AFM) and the effective temperature $T_\text{eff}$ of the Al evaporation which we expect to be close the room temperature at which sample was fabricated. Both Al and Si lattices have cubic symmetry and therefore their stiffness tensor $C$ can be simplified as follows:

\begin{center}
\begin{equation*}
C = 
\begin{pmatrix}
C_{11} & C_{12} & C_{12}  & 0 & 0 & 0  \\
C_{12} & C_{11} & C_{12}  & 0 & 0 & 0  \\
C_{12} & C_{12} & C_{11}  & 0 & 0 & 0  \\
0 & 0 & 0 & C_{44} & 0 & 0 \\
0 & 0 & 0 & 0 & C_{44} & 0 \\
0 & 0 & 0 & 0 & 0 & C_{44}  \\
\end{pmatrix}
\end{equation*}
\end{center}
with $C_{11} = 166~$GPa, $C_{12}= 64~$GPa and $C_{44}= 80~$GPa for anisotropic Si. Al is however modeled as an isotropic material, and its stiffness can be characterized by a Young's modulus of 70~GPa and a Poisson's ratio of 0.33. Material parameters are summarized in Table~\ref{table2}.

\begin{table}[t!]
\caption{Material properties}
\label{table2}
\begin{center}
  \begin{tabular}{ | c | c | c| }
    \hline
     & Aluminum & Silicon  \\ 
     \hline
    Thermal expansion coefficient (1/K) &  $14.3 \times 10^{-6}$ & $0.7 \times 10^{-6}$ \\ \hline
    Stiffness tensor $C_{11},~C_{12},~C_{44}$ (GPa) & $103,~51,~26$ & $166,~64,~80$   \\ \hline
  
  \end{tabular}
\end{center}
\end{table}

\section{Schottky barrier}
\label{app:schottky}
In this paragraph, we discuss the characteristics of the aluminum/silicon interface, and more precisely the potential formation of a Schottky barrier. 
From electrostatic calculation using the known work function of both silicon and aluminum, we would expect formation of a Schottky barrier, with a depletion zone of depth $\sim 100$~\,nm for the donor density in our sample. Therefore, a significant fraction of the donors would be ionized, and particularly those that are closest to the wire edge, which would have a clear spectroscopic signature. On the other hand, Schottky barriers are notoriously difficult to predict, as the presence of additional charged impurities at the silicon/silicon oxide/aluminum interface can modify the barrier properties. 

\begin{figure}[t!]
  \includegraphics[width=\columnwidth]{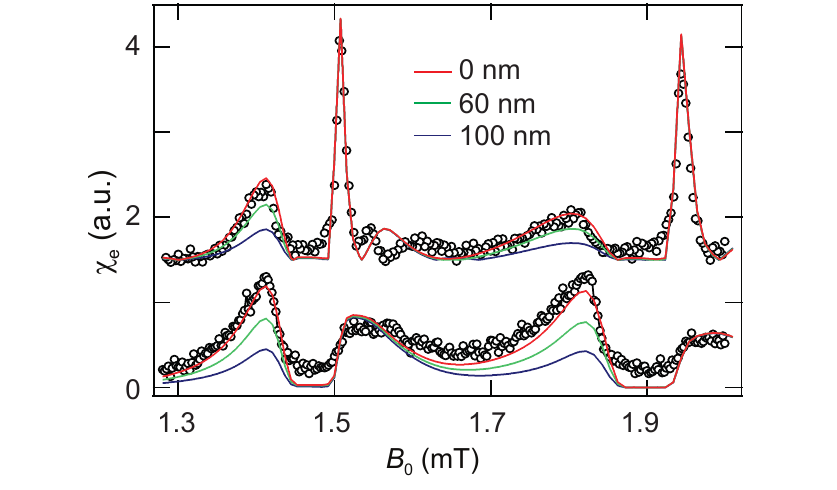}
  \caption{\label{fig:Schottky}
    Simulated spin spectrum (lines) for Res1 when incorporating Schottky barriers of height 0~eV (0~nm), 0.3~eV (60~nm) and 0.7~eV (100~nm). Measured data (symbols) at two pulse amplitudes are the same as shown in the Fig.~\ref{fig:Spectrum}(a). }
\end{figure}

We thus use our spectroscopic data as a way to test three Schottky barrier scenarios: a depletion depth of $100$\,nm ($0.7$\,eV barrier height), $60$\,nm ($0.3$\,eV), and no depletion zone at all. The computed spectra with depletion zones are shown in Fig.~\ref{fig:Schottky} with the same measured experimental data as in Fig.~\ref{fig:Spectrum}(a). The comparison shows better agreement for the scenario without any Schottky barrier, which is the only one reproducing quantitatively the relative height of the low- and high-field peaks. Indeed the shape and dynamics of the Schottky barrier may be significantly affected by the fact that the aluminium strip and silicon are not connected through a circuit to a common reference of chemical potential. The only way Al and Si can exchange charges is through the SiO$_2$ oxide, which can be very slow.

\section{Dynamical decoupling }
    \label{app:DD}

\begin{figure}[t!]
  \includegraphics[width=\columnwidth]{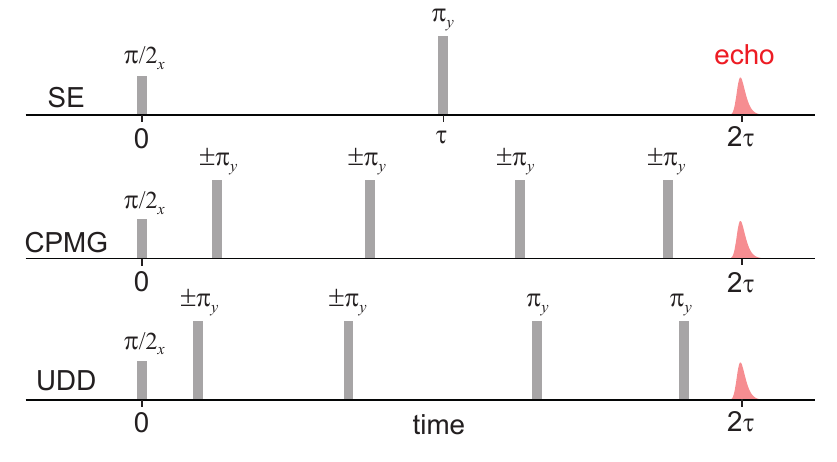}
  \caption{\label{fig:pulseseq}
    Pulse sequences for spin echo, CPMG and UDD ($N=4$). Phase cycling is done by controlling the sign of the $\pi$ pulses. Subscript $x,y$ denote the axis of the pulses.}
\end{figure}

Here we provide more details on the two dynamical decoupling pulse sequences  used in Section~\ref{sec:DDandTemp}.

The Carr-Purcell-Meiboom-Gill (CPMG) sequence consists of a first $\pi/2_x$ pulse, followed after a duration $\tau/N$ by a sequence of $N$ $\pi_y$ pulses separated by $2\tau / N$, generating an echo at time $2\tau$ as seen in Fig.~\ref{fig:pulseseq}. In the Uhrig Dynamical Decoupling (UDD) sequence, the $\pi$ pulses are shifted in time compared to the CPMG, and occur instead at $2\tau \sin^2 [k\pi/(2N+2)]$.

In our experiments, the Rabi rotations are not very precise, because of the spatial inhomogeneity of the $B_1$ field. This gives rise to spurious echoes, because of the phenomenon of stimulated echo where two consecutive $\pi/2$ pulses store coherence as a polarization comb in the frequency domain, which is restored along $S_x$ by a third $\pi/2$ pulse. Such stimulated echoes have a different dependence on the drive pulse phases than the desired echo, resulting from the periodic refocusing of all control pulses. Therefore, phase cycling is the method of choice to measure only the desired echo. 

In the CPMG sequence, because of the equidistance of the $\pi$ pulses, many pulse combinations give rise to a stimulated echo at time $2\tau$. Therefore, a complex phase cycling scheme must be used, varying the phase of each $\pi$ pulse in the sequence (i.e., $2^N$ measurements). The UDD sequence is more favorable, since most stimulated echoes do not happen at $2\tau$ and have therefore no overlap with the desired signal. Only 2 phase cycling sequences are therefore needed for UDD (all $\pi$ pulses in the first-half of the sequence simultaneously take plus or minus sign). The pulse sequences used in this work are shown schematically in Fig.~\ref{fig:pulseseq}.

\section{Meissner Screening}
    \label{app:Meissner}

\begin{figure}[t!]
  \includegraphics[width=\columnwidth]{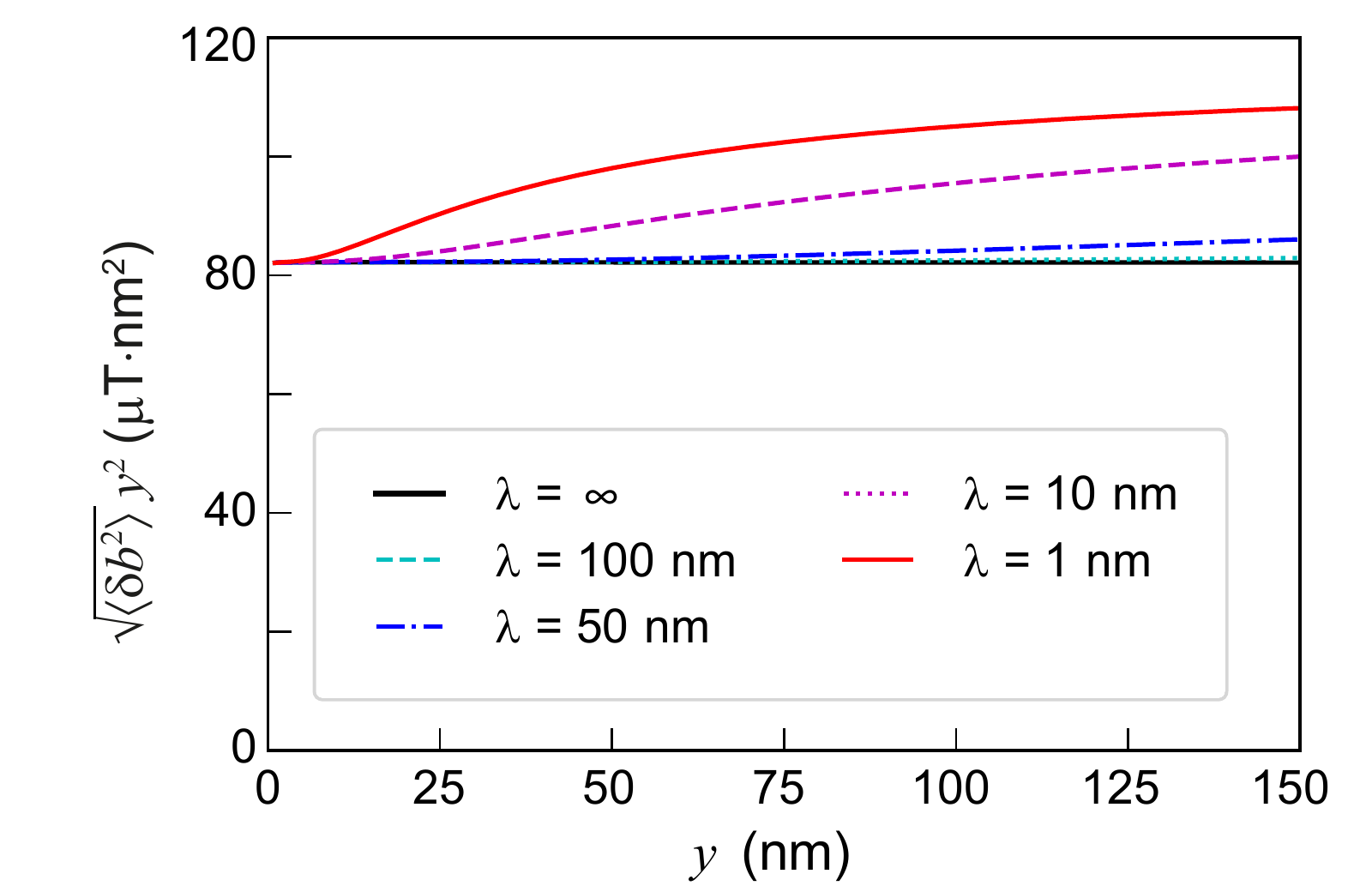}
  \caption{\label{fig:Meissner}
Calculated Meissner screening of magnetic noise due to the superconducting aluminum for different London penetration depths $\lambda$. The magnetic field standard deviations along the $z$-axis, $\sqrt{\langle \delta b^2 \rangle}$ are plotted for randomly oriented surface magnetic impurities with a uniform density of $\sigma = 10^{12}~\text{cm}^{-2}$ and individual magnetic moment of $\mu_\text{B}$.}
\end{figure}  

In order to assess the impact of Meissner effect on magnetic noise, we have computed the screened magnetic field created by a dipole $|\vec{m}|=\mu_\text{B}$ at the Si/SiO$_2$ interface. We assume a 50 nm thick  type I superconducting film ($w\to\infty$), separated from a Si substrate by a 5 nm thick oxide. The screened magnetic field is calculated for a given London penetration depth $\lambda$ as a Fourier-Bessel series solution of Maxwell's equations \cite{milosevic_magnetic_2002}. The squared magnetic field component along $z$, that drives spin decoherence \cite{de_sousa_dangling-bond_2007}, is then averaged over all dipole orientations and over a homogeneous distribution of paramagnetic centers at the Si/SiO$_2$ interface with a density of $\sigma=10^{12}$~ cm$^{-2}$. In the absence of Meissner screening ($\lambda\to\infty$), the resulting $\sqrt{\langle \delta b^2\rangle}$ is expected to decrease as $1/y^2$ in the silicon substrate, with $y$ the distance to the Si/SiO$_2$ interface \cite{de_sousa_dangling-bond_2007}. Therefore, $\sqrt{\langle \delta b^2\rangle} \times y^2$ is plotted for different $\lambda$'s in Fig. \ref{fig:Meissner}. The currents induced in the superconductor do screen the dipoles within that material; they also screen perpendicular dipoles within the semiconductor ($\vec{m}\parallel\vec{y}$), but enhance parallel dipoles ($\vec{m}\parallel\vec{x},\,\vec{z}$). On average over dipole orientations, Meissner screening results in a net enhancement of $\sqrt{\langle \delta b^2\rangle}$ in the semiconductor. For London penetration depths $\lambda>50$ nm, relevant for Aluminium, this enhancement remains moderate and cannot explain the experimental data (it would in fact reduce coherence times under the Al strip). Meissner screening is not, therefore, a leading mechanism in the present experiments.

\section*{Acknowledgements} 
We acknowledge technical support from P.~S\'enat, P.-F.~Orfila and S.~Delprat, and are grateful for fruitful discussions within the Quantronics group. We acknowledge IARPA and Lincoln Labs for providing a JTWPA used in the measurements. We acknowledge support from the Horizon 2020 research and innovation program through grant agreement No. 771493 (LOQO-MOTIONS), and of the European Union through the Marie 365 Sklodowska Curie Grant Agreement No. 765267 (QuSCO), and of the Agence Nationale de la Recherche (ANR) through projects QIPSE, MIRESPIN (ANR 19 CE47 0011) and the Chaire Industrielle NASNIQ  under contract ANR-17-CHIN-0001 cofunded by Atos. T. S. was supported by the U.S. Department of Energy under Contract No. DE-AC02-05CH11231. J.~J.~P. was supported by an Australian Research Council Discovery Early Career Research Award (DE190101397).


\begin{thebibliography}{55}%
\makeatletter
\providecommand \@ifxundefined [1]{%
 \@ifx{#1\undefined}
}%
\providecommand \@ifnum [1]{%
 \ifnum #1\expandafter \@firstoftwo
 \else \expandafter \@secondoftwo
 \fi
}%
\providecommand \@ifx [1]{%
 \ifx #1\expandafter \@firstoftwo
 \else \expandafter \@secondoftwo
 \fi
}%
\providecommand \natexlab [1]{#1}%
\providecommand \enquote  [1]{``#1''}%
\providecommand \bibnamefont  [1]{#1}%
\providecommand \bibfnamefont [1]{#1}%
\providecommand \citenamefont [1]{#1}%
\providecommand \href@noop [0]{\@secondoftwo}%
\providecommand \href [0]{\begingroup \@sanitize@url \@href}%
\providecommand \@href[1]{\@@startlink{#1}\@@href}%
\providecommand \@@href[1]{\endgroup#1\@@endlink}%
\providecommand \@sanitize@url [0]{\catcode `\\12\catcode `\$12\catcode
  `\&12\catcode `\#12\catcode `\^12\catcode `\_12\catcode `\%12\relax}%
\providecommand \@@startlink[1]{}%
\providecommand \@@endlink[0]{}%
\providecommand \url  [0]{\begingroup\@sanitize@url \@url }%
\providecommand \@url [1]{\endgroup\@href {#1}{\urlprefix }}%
\providecommand \urlprefix  [0]{URL }%
\providecommand \Eprint [0]{\href }%
\providecommand \doibase [0]{http://dx.doi.org/}%
\providecommand \selectlanguage [0]{\@gobble}%
\providecommand \bibinfo  [0]{\@secondoftwo}%
\providecommand \bibfield  [0]{\@secondoftwo}%
\providecommand \translation [1]{[#1]}%
\providecommand \BibitemOpen [0]{}%
\providecommand \bibitemStop [0]{}%
\providecommand \bibitemNoStop [0]{.\EOS\space}%
\providecommand \EOS [0]{\spacefactor3000\relax}%
\providecommand \BibitemShut  [1]{\csname bibitem#1\endcsname}%
\let\auto@bib@innerbib\@empty
\bibitem [{\citenamefont {Kane}(1998)}]{kane_silicon-based_1998}%
  \BibitemOpen
  \bibfield  {author} {\bibinfo {author} {\bibfnamefont {B.~E.}\ \bibnamefont
  {Kane}},\ }\href {\doibase 10.1038/30156} {\bibfield  {journal} {\bibinfo
  {journal} {Nature}\ }\textbf {\bibinfo {volume} {393}},\ \bibinfo {pages}
  {133} (\bibinfo {year} {1998})}\BibitemShut {NoStop}%
\bibitem [{\citenamefont {Morello}\ \emph {et~al.}(2010)\citenamefont
  {Morello}, \citenamefont {Pla}, \citenamefont {Zwanenburg}, \citenamefont
  {Chan}, \citenamefont {Tan}, \citenamefont {Huebl}, \citenamefont
  {M{\"o}tt{\"o}nen}, \citenamefont {Nugroho}, \citenamefont {Yang},
  \citenamefont {van Donkelaar}, \citenamefont {Alves}, \citenamefont
  {Jamieson}, \citenamefont {Escott}, \citenamefont {Hollenberg}, \citenamefont
  {Clark},\ and\ \citenamefont {Dzurak}}]{morello_single-shot_2010}%
  \BibitemOpen
  \bibfield  {author} {\bibinfo {author} {\bibfnamefont {A.}~\bibnamefont
  {Morello}}, \bibinfo {author} {\bibfnamefont {J.~J.}\ \bibnamefont {Pla}},
  \bibinfo {author} {\bibfnamefont {F.~A.}\ \bibnamefont {Zwanenburg}},
  \bibinfo {author} {\bibfnamefont {K.~W.}\ \bibnamefont {Chan}}, \bibinfo
  {author} {\bibfnamefont {K.~Y.}\ \bibnamefont {Tan}}, \bibinfo {author}
  {\bibfnamefont {H.}~\bibnamefont {Huebl}}, \bibinfo {author} {\bibfnamefont
  {M.}~\bibnamefont {M{\"o}tt{\"o}nen}}, \bibinfo {author} {\bibfnamefont
  {C.~D.}\ \bibnamefont {Nugroho}}, \bibinfo {author} {\bibfnamefont
  {C.}~\bibnamefont {Yang}}, \bibinfo {author} {\bibfnamefont {J.~A.}\
  \bibnamefont {van Donkelaar}}, \bibinfo {author} {\bibfnamefont {A.~D.~C.}\
  \bibnamefont {Alves}}, \bibinfo {author} {\bibfnamefont {D.~N.}\ \bibnamefont
  {Jamieson}}, \bibinfo {author} {\bibfnamefont {C.~C.}\ \bibnamefont
  {Escott}}, \bibinfo {author} {\bibfnamefont {L.~C.~L.}\ \bibnamefont
  {Hollenberg}}, \bibinfo {author} {\bibfnamefont {R.~G.}\ \bibnamefont
  {Clark}}, \ and\ \bibinfo {author} {\bibfnamefont {A.~S.}\ \bibnamefont
  {Dzurak}},\ }\href {\doibase 10.1038/nature09392} {\bibfield  {journal}
  {\bibinfo  {journal} {Nature}\ }\textbf {\bibinfo {volume} {467}},\ \bibinfo
  {pages} {687} (\bibinfo {year} {2010})}\BibitemShut {NoStop}%
\bibitem [{\citenamefont {Pla}\ \emph {et~al.}(2012)\citenamefont {Pla},
  \citenamefont {Tan}, \citenamefont {Dehollain}, \citenamefont {Lim},
  \citenamefont {Morton}, \citenamefont {Jamieson}, \citenamefont {Dzurak},\
  and\ \citenamefont {Morello}}]{pla_single-atom_2012}%
  \BibitemOpen
  \bibfield  {author} {\bibinfo {author} {\bibfnamefont {J.~J.}\ \bibnamefont
  {Pla}}, \bibinfo {author} {\bibfnamefont {K.~Y.}\ \bibnamefont {Tan}},
  \bibinfo {author} {\bibfnamefont {J.~P.}\ \bibnamefont {Dehollain}}, \bibinfo
  {author} {\bibfnamefont {W.~H.}\ \bibnamefont {Lim}}, \bibinfo {author}
  {\bibfnamefont {J.~J.~L.}\ \bibnamefont {Morton}}, \bibinfo {author}
  {\bibfnamefont {D.~N.}\ \bibnamefont {Jamieson}}, \bibinfo {author}
  {\bibfnamefont {A.~S.}\ \bibnamefont {Dzurak}}, \ and\ \bibinfo {author}
  {\bibfnamefont {A.}~\bibnamefont {Morello}},\ }\href {\doibase
  10.1038/nature11449} {\bibfield  {journal} {\bibinfo  {journal} {Nature}\
  }\textbf {\bibinfo {volume} {489}},\ \bibinfo {pages} {541} (\bibinfo {year}
  {2012})}\BibitemShut {NoStop}%
\bibitem [{\citenamefont {Veldhorst}\ \emph {et~al.}(2014)\citenamefont
  {Veldhorst}, \citenamefont {Hwang}, \citenamefont {Yang}, \citenamefont
  {Leenstra}, \citenamefont {de~Ronde}, \citenamefont {Dehollain},
  \citenamefont {Muhonen}, \citenamefont {Hudson}, \citenamefont {Itoh},
  \citenamefont {Morello},\ and\ \citenamefont
  {Dzurak}}]{veldhorst_addressable_2014}%
  \BibitemOpen
  \bibfield  {author} {\bibinfo {author} {\bibfnamefont {M.}~\bibnamefont
  {Veldhorst}}, \bibinfo {author} {\bibfnamefont {J.~C.~C.}\ \bibnamefont
  {Hwang}}, \bibinfo {author} {\bibfnamefont {C.~H.}\ \bibnamefont {Yang}},
  \bibinfo {author} {\bibfnamefont {A.~W.}\ \bibnamefont {Leenstra}}, \bibinfo
  {author} {\bibfnamefont {B.}~\bibnamefont {de~Ronde}}, \bibinfo {author}
  {\bibfnamefont {J.~P.}\ \bibnamefont {Dehollain}}, \bibinfo {author}
  {\bibfnamefont {J.~T.}\ \bibnamefont {Muhonen}}, \bibinfo {author}
  {\bibfnamefont {F.~E.}\ \bibnamefont {Hudson}}, \bibinfo {author}
  {\bibfnamefont {K.~M.}\ \bibnamefont {Itoh}}, \bibinfo {author}
  {\bibfnamefont {A.}~\bibnamefont {Morello}}, \ and\ \bibinfo {author}
  {\bibfnamefont {A.~S.}\ \bibnamefont {Dzurak}},\ }\href {\doibase
  10.1038/nnano.2014.216} {\bibfield  {journal} {\bibinfo  {journal} {Nature
  Nanotechnology}\ }\textbf {\bibinfo {volume} {9}},\ \bibinfo {pages} {981}
  (\bibinfo {year} {2014})}\BibitemShut {NoStop}%
\bibitem [{\citenamefont {Muhonen}\ \emph {et~al.}(2014)\citenamefont
  {Muhonen}, \citenamefont {Dehollain}, \citenamefont {Laucht}, \citenamefont
  {Hudson}, \citenamefont {Kalra}, \citenamefont {Sekiguchi}, \citenamefont
  {Itoh}, \citenamefont {Jamieson}, \citenamefont {McCallum}, \citenamefont
  {Dzurak},\ and\ \citenamefont {Morello}}]{muhonen_storing_2014}%
  \BibitemOpen
  \bibfield  {author} {\bibinfo {author} {\bibfnamefont {J.~T.}\ \bibnamefont
  {Muhonen}}, \bibinfo {author} {\bibfnamefont {J.~P.}\ \bibnamefont
  {Dehollain}}, \bibinfo {author} {\bibfnamefont {A.}~\bibnamefont {Laucht}},
  \bibinfo {author} {\bibfnamefont {F.~E.}\ \bibnamefont {Hudson}}, \bibinfo
  {author} {\bibfnamefont {R.}~\bibnamefont {Kalra}}, \bibinfo {author}
  {\bibfnamefont {T.}~\bibnamefont {Sekiguchi}}, \bibinfo {author}
  {\bibfnamefont {K.~M.}\ \bibnamefont {Itoh}}, \bibinfo {author}
  {\bibfnamefont {D.~N.}\ \bibnamefont {Jamieson}}, \bibinfo {author}
  {\bibfnamefont {J.~C.}\ \bibnamefont {McCallum}}, \bibinfo {author}
  {\bibfnamefont {A.~S.}\ \bibnamefont {Dzurak}}, \ and\ \bibinfo {author}
  {\bibfnamefont {A.}~\bibnamefont {Morello}},\ }\href {\doibase
  10.1038/nnano.2014.211} {\bibfield  {journal} {\bibinfo  {journal} {Nature
  Nanotechnology}\ }\textbf {\bibinfo {volume} {9}},\ \bibinfo {pages} {986}
  (\bibinfo {year} {2014})}\BibitemShut {NoStop}%
\bibitem [{\citenamefont {Yang}\ \emph {et~al.}(2020)\citenamefont {Yang},
  \citenamefont {Leon}, \citenamefont {Hwang}, \citenamefont {Saraiva},
  \citenamefont {Tanttu}, \citenamefont {Huang}, \citenamefont
  {Camirand~Lemyre}, \citenamefont {Chan}, \citenamefont {Tan}, \citenamefont
  {Hudson}, \citenamefont {Itoh}, \citenamefont {Morello}, \citenamefont
  {Pioro-Ladri{\`e}re}, \citenamefont {Laucht},\ and\ \citenamefont
  {Dzurak}}]{yang_operation_2020}%
  \BibitemOpen
  \bibfield  {author} {\bibinfo {author} {\bibfnamefont {C.~H.}\ \bibnamefont
  {Yang}}, \bibinfo {author} {\bibfnamefont {R.~C.~C.}\ \bibnamefont {Leon}},
  \bibinfo {author} {\bibfnamefont {J.~C.~C.}\ \bibnamefont {Hwang}}, \bibinfo
  {author} {\bibfnamefont {A.}~\bibnamefont {Saraiva}}, \bibinfo {author}
  {\bibfnamefont {T.}~\bibnamefont {Tanttu}}, \bibinfo {author} {\bibfnamefont
  {W.}~\bibnamefont {Huang}}, \bibinfo {author} {\bibfnamefont
  {J.}~\bibnamefont {Camirand~Lemyre}}, \bibinfo {author} {\bibfnamefont
  {K.~W.}\ \bibnamefont {Chan}}, \bibinfo {author} {\bibfnamefont {K.~Y.}\
  \bibnamefont {Tan}}, \bibinfo {author} {\bibfnamefont {F.~E.}\ \bibnamefont
  {Hudson}}, \bibinfo {author} {\bibfnamefont {K.~M.}\ \bibnamefont {Itoh}},
  \bibinfo {author} {\bibfnamefont {A.}~\bibnamefont {Morello}}, \bibinfo
  {author} {\bibfnamefont {M.}~\bibnamefont {Pioro-Ladri{\`e}re}}, \bibinfo
  {author} {\bibfnamefont {A.}~\bibnamefont {Laucht}}, \ and\ \bibinfo {author}
  {\bibfnamefont {A.~S.}\ \bibnamefont {Dzurak}},\ }\href {\doibase
  10.1038/s41586-020-2171-6} {\bibfield  {journal} {\bibinfo  {journal}
  {Nature}\ }\textbf {\bibinfo {volume} {580}},\ \bibinfo {pages} {350}
  (\bibinfo {year} {2020})}\BibitemShut {NoStop}%
\bibitem [{\citenamefont {Petit}\ \emph {et~al.}(2020)\citenamefont {Petit},
  \citenamefont {Eenink}, \citenamefont {Russ}, \citenamefont {Lawrie},
  \citenamefont {Hendrickx}, \citenamefont {Philips}, \citenamefont {Clarke},
  \citenamefont {Vandersypen},\ and\ \citenamefont
  {Veldhorst}}]{petit_universal_2020}%
  \BibitemOpen
  \bibfield  {author} {\bibinfo {author} {\bibfnamefont {L.}~\bibnamefont
  {Petit}}, \bibinfo {author} {\bibfnamefont {H.~G.~J.}\ \bibnamefont
  {Eenink}}, \bibinfo {author} {\bibfnamefont {M.}~\bibnamefont {Russ}},
  \bibinfo {author} {\bibfnamefont {W.~I.~L.}\ \bibnamefont {Lawrie}}, \bibinfo
  {author} {\bibfnamefont {N.~W.}\ \bibnamefont {Hendrickx}}, \bibinfo {author}
  {\bibfnamefont {S.~G.~J.}\ \bibnamefont {Philips}}, \bibinfo {author}
  {\bibfnamefont {J.~S.}\ \bibnamefont {Clarke}}, \bibinfo {author}
  {\bibfnamefont {L.~M.~K.}\ \bibnamefont {Vandersypen}}, \ and\ \bibinfo
  {author} {\bibfnamefont {M.}~\bibnamefont {Veldhorst}},\ }\href {\doibase
  10.1038/s41586-020-2170-7} {\bibfield  {journal} {\bibinfo  {journal}
  {Nature}\ }\textbf {\bibinfo {volume} {580}},\ \bibinfo {pages} {355}
  (\bibinfo {year} {2020})}\BibitemShut {NoStop}%
\bibitem [{\citenamefont {Morello}\ \emph {et~al.}()\citenamefont {Morello},
  \citenamefont {Pla}, \citenamefont {Bertet},\ and\ \citenamefont
  {Jamieson}}]{morello_donor_nodate}%
  \BibitemOpen
  \bibfield  {author} {\bibinfo {author} {\bibfnamefont {A.}~\bibnamefont
  {Morello}}, \bibinfo {author} {\bibfnamefont {J.~J.}\ \bibnamefont {Pla}},
  \bibinfo {author} {\bibfnamefont {P.}~\bibnamefont {Bertet}}, \ and\ \bibinfo
  {author} {\bibfnamefont {D.~N.}\ \bibnamefont {Jamieson}},\ }\href {\doibase
  10.1002/qute.202000005} {\bibfield  {journal} {\bibinfo  {journal} {Advanced
  Quantum Technologies}\ }\textbf {\bibinfo {volume} {n/a}},\ \bibinfo {pages}
  {2000005}}\BibitemShut {NoStop}%
\bibitem [{\citenamefont {Tyryshkin}\ \emph {et~al.}(2012)\citenamefont
  {Tyryshkin}, \citenamefont {Tojo}, \citenamefont {Morton}, \citenamefont
  {Riemann}, \citenamefont {Abrosimov}, \citenamefont {Becker}, \citenamefont
  {Pohl}, \citenamefont {Schenkel}, \citenamefont {Thewalt}, \citenamefont
  {Itoh},\ and\ \citenamefont {Lyon}}]{tyryshkin_electron_2012}%
  \BibitemOpen
  \bibfield  {author} {\bibinfo {author} {\bibfnamefont {A.~M.}\ \bibnamefont
  {Tyryshkin}}, \bibinfo {author} {\bibfnamefont {S.}~\bibnamefont {Tojo}},
  \bibinfo {author} {\bibfnamefont {J.~J.~L.}\ \bibnamefont {Morton}}, \bibinfo
  {author} {\bibfnamefont {H.}~\bibnamefont {Riemann}}, \bibinfo {author}
  {\bibfnamefont {N.~V.}\ \bibnamefont {Abrosimov}}, \bibinfo {author}
  {\bibfnamefont {P.}~\bibnamefont {Becker}}, \bibinfo {author} {\bibfnamefont
  {H.-J.}\ \bibnamefont {Pohl}}, \bibinfo {author} {\bibfnamefont
  {T.}~\bibnamefont {Schenkel}}, \bibinfo {author} {\bibfnamefont {M.~L.~W.}\
  \bibnamefont {Thewalt}}, \bibinfo {author} {\bibfnamefont {K.~M.}\
  \bibnamefont {Itoh}}, \ and\ \bibinfo {author} {\bibfnamefont {S.~A.}\
  \bibnamefont {Lyon}},\ }\href {\doibase 10.1038/nmat3182} {\bibfield
  {journal} {\bibinfo  {journal} {Nature Materials}\ }\textbf {\bibinfo
  {volume} {11}},\ \bibinfo {pages} {143} (\bibinfo {year} {2012})}\BibitemShut
  {NoStop}%
\bibitem [{\citenamefont {Wolfowicz}\ \emph {et~al.}(2013)\citenamefont
  {Wolfowicz}, \citenamefont {Tyryshkin}, \citenamefont {George}, \citenamefont
  {Riemann}, \citenamefont {Abrosimov}, \citenamefont {Becker}, \citenamefont
  {Pohl}, \citenamefont {Thewalt}, \citenamefont {Lyon},\ and\ \citenamefont
  {Morton}}]{wolfowicz_atomic_2013}%
  \BibitemOpen
  \bibfield  {author} {\bibinfo {author} {\bibfnamefont {G.}~\bibnamefont
  {Wolfowicz}}, \bibinfo {author} {\bibfnamefont {A.~M.}\ \bibnamefont
  {Tyryshkin}}, \bibinfo {author} {\bibfnamefont {R.~E.}\ \bibnamefont
  {George}}, \bibinfo {author} {\bibfnamefont {H.}~\bibnamefont {Riemann}},
  \bibinfo {author} {\bibfnamefont {N.~V.}\ \bibnamefont {Abrosimov}}, \bibinfo
  {author} {\bibfnamefont {P.}~\bibnamefont {Becker}}, \bibinfo {author}
  {\bibfnamefont {H.-J.}\ \bibnamefont {Pohl}}, \bibinfo {author}
  {\bibfnamefont {M.~L.~W.}\ \bibnamefont {Thewalt}}, \bibinfo {author}
  {\bibfnamefont {S.~a.}\ \bibnamefont {Lyon}}, \ and\ \bibinfo {author}
  {\bibfnamefont {J.~J.~L.}\ \bibnamefont {Morton}},\ }\href {\doibase
  10.1038/nnano.2013.117} {\bibfield  {journal} {\bibinfo  {journal} {Nature
  Nanotechnology}\ }\textbf {\bibinfo {volume} {8}},\ \bibinfo {pages} {561}
  (\bibinfo {year} {2013})}\BibitemShut {NoStop}%
\bibitem [{\citenamefont {Julsgaard}\ \emph {et~al.}(2013)\citenamefont
  {Julsgaard}, \citenamefont {Grezes}, \citenamefont {Bertet},\ and\
  \citenamefont {Molmer}}]{julsgaard_quantum_2013}%
  \BibitemOpen
  \bibfield  {author} {\bibinfo {author} {\bibfnamefont {B.}~\bibnamefont
  {Julsgaard}}, \bibinfo {author} {\bibfnamefont {C.}~\bibnamefont {Grezes}},
  \bibinfo {author} {\bibfnamefont {P.}~\bibnamefont {Bertet}}, \ and\ \bibinfo
  {author} {\bibfnamefont {K.}~\bibnamefont {Molmer}},\ }\href {\doibase
  10.1103/PhysRevLett.110.250503} {\bibfield  {journal} {\bibinfo  {journal}
  {Physical Review Letters}\ }\textbf {\bibinfo {volume} {110}} (\bibinfo
  {year} {2013}),\ 10.1103/PhysRevLett.110.250503}\BibitemShut {NoStop}%
\bibitem [{\citenamefont {Morton}\ and\ \citenamefont
  {Bertet}(2018)}]{morton_storing_2018}%
  \BibitemOpen
  \bibfield  {author} {\bibinfo {author} {\bibfnamefont {J.~J.~L.}\
  \bibnamefont {Morton}}\ and\ \bibinfo {author} {\bibfnamefont
  {P.}~\bibnamefont {Bertet}},\ }\href {\doibase 10.1016/j.jmr.2017.11.015}
  {\bibfield  {journal} {\bibinfo  {journal} {Journal of Magnetic Resonance}\
  }\textbf {\bibinfo {volume} {287}},\ \bibinfo {pages} {128} (\bibinfo {year}
  {2018})}\BibitemShut {NoStop}%
\bibitem [{\citenamefont {Ranjan}\ \emph
  {et~al.}(2020{\natexlab{a}})\citenamefont {Ranjan}, \citenamefont
  {O{\textquoteright}Sullivan}, \citenamefont {Albertinale}, \citenamefont
  {Albanese}, \citenamefont {Chaneli{\`e}re}, \citenamefont {Schenkel},
  \citenamefont {Vion}, \citenamefont {Esteve}, \citenamefont {Flurin},
  \citenamefont {Morton},\ and\ \citenamefont
  {Bertet}}]{ranjan_multimode_2020-1}%
  \BibitemOpen
  \bibfield  {author} {\bibinfo {author} {\bibfnamefont {V.}~\bibnamefont
  {Ranjan}}, \bibinfo {author} {\bibfnamefont {J.}~\bibnamefont
  {O{\textquoteright}Sullivan}}, \bibinfo {author} {\bibfnamefont
  {E.}~\bibnamefont {Albertinale}}, \bibinfo {author} {\bibfnamefont
  {B.}~\bibnamefont {Albanese}}, \bibinfo {author} {\bibfnamefont
  {T.}~\bibnamefont {Chaneli{\`e}re}}, \bibinfo {author} {\bibfnamefont
  {T.}~\bibnamefont {Schenkel}}, \bibinfo {author} {\bibfnamefont
  {D.}~\bibnamefont {Vion}}, \bibinfo {author} {\bibfnamefont {D.}~\bibnamefont
  {Esteve}}, \bibinfo {author} {\bibfnamefont {E.}~\bibnamefont {Flurin}},
  \bibinfo {author} {\bibfnamefont {J.~J.~L.}\ \bibnamefont {Morton}}, \ and\
  \bibinfo {author} {\bibfnamefont {P.}~\bibnamefont {Bertet}},\ }\href
  {\doibase 10.1103/PhysRevLett.125.210505} {\bibfield  {journal} {\bibinfo
  {journal} {Physical Review Letters}\ }\textbf {\bibinfo {volume} {125}},\
  \bibinfo {pages} {210505} (\bibinfo {year} {2020}{\natexlab{a}})},\ \bibinfo
  {note} {publisher: American Physical Society}\BibitemShut {NoStop}%
\bibitem [{\citenamefont {Schenkel}\ \emph {et~al.}(2006)\citenamefont
  {Schenkel}, \citenamefont {Liddle}, \citenamefont {Persaud}, \citenamefont
  {Tyryshkin}, \citenamefont {Lyon}, \citenamefont {de~Sousa}, \citenamefont
  {Whaley}, \citenamefont {Bokor}, \citenamefont {Shangkuan},\ and\
  \citenamefont {Chakarov}}]{schenkel_electrical_2006}%
  \BibitemOpen
  \bibfield  {author} {\bibinfo {author} {\bibfnamefont {T.}~\bibnamefont
  {Schenkel}}, \bibinfo {author} {\bibfnamefont {J.~A.}\ \bibnamefont
  {Liddle}}, \bibinfo {author} {\bibfnamefont {A.}~\bibnamefont {Persaud}},
  \bibinfo {author} {\bibfnamefont {A.~M.}\ \bibnamefont {Tyryshkin}}, \bibinfo
  {author} {\bibfnamefont {S.~A.}\ \bibnamefont {Lyon}}, \bibinfo {author}
  {\bibfnamefont {R.}~\bibnamefont {de~Sousa}}, \bibinfo {author}
  {\bibfnamefont {K.~B.}\ \bibnamefont {Whaley}}, \bibinfo {author}
  {\bibfnamefont {J.}~\bibnamefont {Bokor}}, \bibinfo {author} {\bibfnamefont
  {J.}~\bibnamefont {Shangkuan}}, \ and\ \bibinfo {author} {\bibfnamefont
  {I.}~\bibnamefont {Chakarov}},\ }\href {\doibase 10.1063/1.2182068}
  {\bibfield  {journal} {\bibinfo  {journal} {Applied Physics Letters}\
  }\textbf {\bibinfo {volume} {88}},\ \bibinfo {pages} {112101} (\bibinfo
  {year} {2006})}\BibitemShut {NoStop}%
\bibitem [{\citenamefont {de~Sousa}(2007)}]{de_sousa_dangling-bond_2007}%
  \BibitemOpen
  \bibfield  {author} {\bibinfo {author} {\bibfnamefont {R.}~\bibnamefont
  {de~Sousa}},\ }\href {\doibase 10.1103/PhysRevB.76.245306} {\bibfield
  {journal} {\bibinfo  {journal} {Physical Review B}\ }\textbf {\bibinfo
  {volume} {76}},\ \bibinfo {pages} {245306} (\bibinfo {year}
  {2007})}\BibitemShut {NoStop}%
\bibitem [{\citenamefont {Paik}\ \emph {et~al.}(2010)\citenamefont {Paik},
  \citenamefont {Lee}, \citenamefont {Baker}, \citenamefont {McCamey},\ and\
  \citenamefont {Boehme}}]{paik_t_1_2010}%
  \BibitemOpen
  \bibfield  {author} {\bibinfo {author} {\bibfnamefont {S.-Y.}\ \bibnamefont
  {Paik}}, \bibinfo {author} {\bibfnamefont {S.-Y.}\ \bibnamefont {Lee}},
  \bibinfo {author} {\bibfnamefont {W.~J.}\ \bibnamefont {Baker}}, \bibinfo
  {author} {\bibfnamefont {D.~R.}\ \bibnamefont {McCamey}}, \ and\ \bibinfo
  {author} {\bibfnamefont {C.}~\bibnamefont {Boehme}},\ }\href {\doibase
  10.1103/PhysRevB.81.075214} {\bibfield  {journal} {\bibinfo  {journal}
  {Physical Review B}\ }\textbf {\bibinfo {volume} {81}},\ \bibinfo {pages}
  {075214} (\bibinfo {year} {2010})}\BibitemShut {NoStop}%
\bibitem [{\citenamefont {Tenberg}\ \emph {et~al.}(2019)\citenamefont
  {Tenberg}, \citenamefont {Asaad}, \citenamefont {Mkadzik}, \citenamefont
  {Johnson}, \citenamefont {Joecker}, \citenamefont {Laucht}, \citenamefont
  {Hudson}, \citenamefont {Itoh}, \citenamefont {Jakob}, \citenamefont
  {Johnson}, \citenamefont {Jamieson}, \citenamefont {McCallum}, \citenamefont
  {Dzurak}, \citenamefont {Joynt},\ and\ \citenamefont
  {Morello}}]{tenberg_electron_2019}%
  \BibitemOpen
  \bibfield  {author} {\bibinfo {author} {\bibfnamefont {S.~B.}\ \bibnamefont
  {Tenberg}}, \bibinfo {author} {\bibfnamefont {S.}~\bibnamefont {Asaad}},
  \bibinfo {author} {\bibfnamefont {M.~T.}\ \bibnamefont {Mkadzik}},
  \bibinfo {author} {\bibfnamefont {M.~A.~I.}\ \bibnamefont {Johnson}},
  \bibinfo {author} {\bibfnamefont {B.}~\bibnamefont {Joecker}}, \bibinfo
  {author} {\bibfnamefont {A.}~\bibnamefont {Laucht}}, \bibinfo {author}
  {\bibfnamefont {F.~E.}\ \bibnamefont {Hudson}}, \bibinfo {author}
  {\bibfnamefont {K.~M.}\ \bibnamefont {Itoh}}, \bibinfo {author}
  {\bibfnamefont {A.~M.}\ \bibnamefont {Jakob}}, \bibinfo {author}
  {\bibfnamefont {B.~C.}\ \bibnamefont {Johnson}}, \bibinfo {author}
  {\bibfnamefont {D.~N.}\ \bibnamefont {Jamieson}}, \bibinfo {author}
  {\bibfnamefont {J.~C.}\ \bibnamefont {McCallum}}, \bibinfo {author}
  {\bibfnamefont {A.~S.}\ \bibnamefont {Dzurak}}, \bibinfo {author}
  {\bibfnamefont {R.}~\bibnamefont {Joynt}}, \ and\ \bibinfo {author}
  {\bibfnamefont {A.}~\bibnamefont {Morello}},\ }\href {\doibase
  10.1103/PhysRevB.99.205306} {\bibfield  {journal} {\bibinfo  {journal}
  {Physical Review B}\ }\textbf {\bibinfo {volume} {99}},\ \bibinfo {pages}
  {205306} (\bibinfo {year} {2019})}\BibitemShut {NoStop}%
\bibitem [{\citenamefont {Mansir}\ \emph {et~al.}(2018)\citenamefont {Mansir},
  \citenamefont {Conti}, \citenamefont {Zeng}, \citenamefont {Pla},
  \citenamefont {Bertet}, \citenamefont {Swift}, \citenamefont {Van~de Walle},
  \citenamefont {Thewalt}, \citenamefont {Sklenard}, \citenamefont {Niquet},\
  and\ \citenamefont {Morton}}]{mansir_linear_2018}%
  \BibitemOpen
  \bibfield  {author} {\bibinfo {author} {\bibfnamefont {J.}~\bibnamefont
  {Mansir}}, \bibinfo {author} {\bibfnamefont {P.}~\bibnamefont {Conti}},
  \bibinfo {author} {\bibfnamefont {Z.}~\bibnamefont {Zeng}}, \bibinfo {author}
  {\bibfnamefont {J.~J.}\ \bibnamefont {Pla}}, \bibinfo {author} {\bibfnamefont
  {P.}~\bibnamefont {Bertet}}, \bibinfo {author} {\bibfnamefont {M.~W.}\
  \bibnamefont {Swift}}, \bibinfo {author} {\bibfnamefont {C.~G.}\ \bibnamefont
  {Van~de Walle}}, \bibinfo {author} {\bibfnamefont {M.~L.~W.}\ \bibnamefont
  {Thewalt}}, \bibinfo {author} {\bibfnamefont {B.}~\bibnamefont {Sklenard}},
  \bibinfo {author} {\bibfnamefont {Y.~M.}\ \bibnamefont {Niquet}}, \ and\
  \bibinfo {author} {\bibfnamefont {J.~J.~L.}\ \bibnamefont {Morton}},\ }\href
  {\doibase 10.1103/PhysRevLett.120.167701} {\bibfield  {journal} {\bibinfo
  {journal} {Physical Review Letters}\ }\textbf {\bibinfo {volume} {120}},\
  \bibinfo {pages} {167701} (\bibinfo {year} {2018})}\BibitemShut {NoStop}%
\bibitem [{\citenamefont {Bienfait}\ \emph
  {et~al.}(2016{\natexlab{a}})\citenamefont {Bienfait}, \citenamefont {Pla},
  \citenamefont {Kubo}, \citenamefont {Stern}, \citenamefont {Zhou},
  \citenamefont {Lo}, \citenamefont {Weis}, \citenamefont {Schenkel},
  \citenamefont {Thewalt}, \citenamefont {Vion}, \citenamefont {Esteve},
  \citenamefont {Julsgaard}, \citenamefont {M{\o}lmer}, \citenamefont
  {Morton},\ and\ \citenamefont {Bertet}}]{bienfait_reaching_2016}%
  \BibitemOpen
  \bibfield  {author} {\bibinfo {author} {\bibfnamefont {A.}~\bibnamefont
  {Bienfait}}, \bibinfo {author} {\bibfnamefont {J.~J.}\ \bibnamefont {Pla}},
  \bibinfo {author} {\bibfnamefont {Y.}~\bibnamefont {Kubo}}, \bibinfo {author}
  {\bibfnamefont {M.}~\bibnamefont {Stern}}, \bibinfo {author} {\bibfnamefont
  {X.}~\bibnamefont {Zhou}}, \bibinfo {author} {\bibfnamefont {C.~C.}\
  \bibnamefont {Lo}}, \bibinfo {author} {\bibfnamefont {C.~D.}\ \bibnamefont
  {Weis}}, \bibinfo {author} {\bibfnamefont {T.}~\bibnamefont {Schenkel}},
  \bibinfo {author} {\bibfnamefont {M.~L.~W.}\ \bibnamefont {Thewalt}},
  \bibinfo {author} {\bibfnamefont {D.}~\bibnamefont {Vion}}, \bibinfo {author}
  {\bibfnamefont {D.}~\bibnamefont {Esteve}}, \bibinfo {author} {\bibfnamefont
  {B.}~\bibnamefont {Julsgaard}}, \bibinfo {author} {\bibfnamefont
  {K.}~\bibnamefont {M{\o}lmer}}, \bibinfo {author} {\bibfnamefont {J.~J.~L.}\
  \bibnamefont {Morton}}, \ and\ \bibinfo {author} {\bibfnamefont
  {P.}~\bibnamefont {Bertet}},\ }\href {\doibase 10.1038/nnano.2015.282}
  {\bibfield  {journal} {\bibinfo  {journal} {Nature Nanotechnology}\ }\textbf
  {\bibinfo {volume} {11}},\ \bibinfo {pages} {253} (\bibinfo {year}
  {2016}{\natexlab{a}})}\BibitemShut {NoStop}%
\bibitem [{\citenamefont {Probst}\ \emph {et~al.}(2017)\citenamefont {Probst},
  \citenamefont {Bienfait}, \citenamefont {Campagne-Ibarcq}, \citenamefont
  {Pla}, \citenamefont {Albanese}, \citenamefont {Barbosa}, \citenamefont
  {Schenkel}, \citenamefont {Vion}, \citenamefont {Esteve}, \citenamefont
  {Moelmer}, \citenamefont {Morton}, \citenamefont {Heeres},\ and\
  \citenamefont {Bertet}}]{probst_inductive-detection_2017}%
  \BibitemOpen
  \bibfield  {author} {\bibinfo {author} {\bibfnamefont {S.}~\bibnamefont
  {Probst}}, \bibinfo {author} {\bibfnamefont {A.}~\bibnamefont {Bienfait}},
  \bibinfo {author} {\bibfnamefont {P.}~\bibnamefont {Campagne-Ibarcq}},
  \bibinfo {author} {\bibfnamefont {J.~J.}\ \bibnamefont {Pla}}, \bibinfo
  {author} {\bibfnamefont {B.}~\bibnamefont {Albanese}}, \bibinfo {author}
  {\bibfnamefont {J.~F. D.~S.}\ \bibnamefont {Barbosa}}, \bibinfo {author}
  {\bibfnamefont {T.}~\bibnamefont {Schenkel}}, \bibinfo {author}
  {\bibfnamefont {D.}~\bibnamefont {Vion}}, \bibinfo {author} {\bibfnamefont
  {D.}~\bibnamefont {Esteve}}, \bibinfo {author} {\bibfnamefont
  {K.}~\bibnamefont {Moelmer}}, \bibinfo {author} {\bibfnamefont {J.~J.~L.}\
  \bibnamefont {Morton}}, \bibinfo {author} {\bibfnamefont {R.}~\bibnamefont
  {Heeres}}, \ and\ \bibinfo {author} {\bibfnamefont {P.}~\bibnamefont
  {Bertet}},\ }\href {\doibase 10.1063/1.5002540} {\bibfield  {journal}
  {\bibinfo  {journal} {Applied Physics Letters}\ }\textbf {\bibinfo {volume}
  {111}},\ \bibinfo {pages} {202604} (\bibinfo {year} {2017})}\BibitemShut
  {NoStop}%
\bibitem [{\citenamefont {Ranjan}\ \emph
  {et~al.}(2020{\natexlab{b}})\citenamefont {Ranjan}, \citenamefont {Probst},
  \citenamefont {Albanese}, \citenamefont {Schenkel}, \citenamefont {Vion},
  \citenamefont {Esteve}, \citenamefont {Morton},\ and\ \citenamefont
  {Bertet}}]{ranjan_electron_2020}%
  \BibitemOpen
  \bibfield  {author} {\bibinfo {author} {\bibfnamefont {V.}~\bibnamefont
  {Ranjan}}, \bibinfo {author} {\bibfnamefont {S.}~\bibnamefont {Probst}},
  \bibinfo {author} {\bibfnamefont {B.}~\bibnamefont {Albanese}}, \bibinfo
  {author} {\bibfnamefont {T.}~\bibnamefont {Schenkel}}, \bibinfo {author}
  {\bibfnamefont {D.}~\bibnamefont {Vion}}, \bibinfo {author} {\bibfnamefont
  {D.}~\bibnamefont {Esteve}}, \bibinfo {author} {\bibfnamefont {J.~J.~L.}\
  \bibnamefont {Morton}}, \ and\ \bibinfo {author} {\bibfnamefont
  {P.}~\bibnamefont {Bertet}},\ }\href {\doibase 10.1063/5.0004322} {\bibfield
  {journal} {\bibinfo  {journal} {Applied Physics Letters}\ }\textbf {\bibinfo
  {volume} {116}},\ \bibinfo {pages} {184002} (\bibinfo {year}
  {2020}{\natexlab{b}})},\ \bibinfo {note} {publisher: American Institute of
  Physics}\BibitemShut {NoStop}%
\bibitem [{\citenamefont {Pla}\ \emph {et~al.}(2018)\citenamefont {Pla},
  \citenamefont {Bienfait}, \citenamefont {Pica}, \citenamefont {Mansir},
  \citenamefont {Mohiyaddin}, \citenamefont {Zeng}, \citenamefont {Niquet},
  \citenamefont {Morello}, \citenamefont {Schenkel}, \citenamefont {Morton},\
  and\ \citenamefont {Bertet}}]{pla_strain-induced_2018}%
  \BibitemOpen
  \bibfield  {author} {\bibinfo {author} {\bibfnamefont {J.~J.}\ \bibnamefont
  {Pla}}, \bibinfo {author} {\bibfnamefont {A.}~\bibnamefont {Bienfait}},
  \bibinfo {author} {\bibfnamefont {G.}~\bibnamefont {Pica}}, \bibinfo {author}
  {\bibfnamefont {J.}~\bibnamefont {Mansir}}, \bibinfo {author} {\bibfnamefont
  {F.~A.}\ \bibnamefont {Mohiyaddin}}, \bibinfo {author} {\bibfnamefont
  {Z.}~\bibnamefont {Zeng}}, \bibinfo {author} {\bibfnamefont {Y.~M.}\
  \bibnamefont {Niquet}}, \bibinfo {author} {\bibfnamefont {A.}~\bibnamefont
  {Morello}}, \bibinfo {author} {\bibfnamefont {T.}~\bibnamefont {Schenkel}},
  \bibinfo {author} {\bibfnamefont {J.~J.~L.}\ \bibnamefont {Morton}}, \ and\
  \bibinfo {author} {\bibfnamefont {P.}~\bibnamefont {Bertet}},\ }\href
  {\doibase 10.1103/PhysRevApplied.9.044014} {\bibfield  {journal} {\bibinfo
  {journal} {Physical Review Applied}\ }\textbf {\bibinfo {volume} {9}},\
  \bibinfo {pages} {044014} (\bibinfo {year} {2018})}\BibitemShut {NoStop}%
\bibitem [{\citenamefont {O{\textquoteright}Sullivan}\ \emph
  {et~al.}(2020)\citenamefont {O{\textquoteright}Sullivan}, \citenamefont
  {Kennedy}, \citenamefont {Zollitsch}, \citenamefont {{\v S}im{\.e}nas},
  \citenamefont {Thomas}, \citenamefont {Abdurakhimov}, \citenamefont
  {Withington},\ and\ \citenamefont {Morton}}]{osullivan_spin-resonance_2020}%
  \BibitemOpen
  \bibfield  {author} {\bibinfo {author} {\bibfnamefont {J.}~\bibnamefont
  {O{\textquoteright}Sullivan}}, \bibinfo {author} {\bibfnamefont {O.~W.}\
  \bibnamefont {Kennedy}}, \bibinfo {author} {\bibfnamefont {C.~W.}\
  \bibnamefont {Zollitsch}}, \bibinfo {author} {\bibfnamefont {M.}~\bibnamefont
  {{\v S}im{\.e}nas}}, \bibinfo {author} {\bibfnamefont {C.~N.}\ \bibnamefont
  {Thomas}}, \bibinfo {author} {\bibfnamefont {L.~V.}\ \bibnamefont
  {Abdurakhimov}}, \bibinfo {author} {\bibfnamefont {S.}~\bibnamefont
  {Withington}}, \ and\ \bibinfo {author} {\bibfnamefont {J.~J.}\ \bibnamefont
  {Morton}},\ }\href {\doibase 10.1103/PhysRevApplied.14.064050} {\bibfield
  {journal} {\bibinfo  {journal} {Physical Review Applied}\ }\textbf {\bibinfo
  {volume} {14}},\ \bibinfo {pages} {064050} (\bibinfo {year} {2020})},\
  \bibinfo {note} {publisher: American Physical Society}\BibitemShut {NoStop}%
\bibitem [{\citenamefont {Feher}(1959)}]{feher_electron_1959}%
  \BibitemOpen
  \bibfield  {author} {\bibinfo {author} {\bibfnamefont {G.}~\bibnamefont
  {Feher}},\ }\href {\doibase 10.1103/PhysRev.114.1219} {\bibfield  {journal}
  {\bibinfo  {journal} {Phys. Rev.}\ }\textbf {\bibinfo {volume} {114}},\
  \bibinfo {pages} {1219} (\bibinfo {year} {1959})}\BibitemShut {NoStop}%
\bibitem [{\citenamefont {Mohammady}\ \emph {et~al.}(2010)\citenamefont
  {Mohammady}, \citenamefont {Morley},\ and\ \citenamefont
  {Monteiro}}]{mohammady_bismuth_2010}%
  \BibitemOpen
  \bibfield  {author} {\bibinfo {author} {\bibfnamefont {M.~H.}\ \bibnamefont
  {Mohammady}}, \bibinfo {author} {\bibfnamefont {G.~W.}\ \bibnamefont
  {Morley}}, \ and\ \bibinfo {author} {\bibfnamefont {T.~S.}\ \bibnamefont
  {Monteiro}},\ }\href {\doibase 10.1103/physrevlett.105.067602} {\bibfield
  {journal} {\bibinfo  {journal} {Physical Review Letters}\ }\textbf {\bibinfo
  {volume} {105}},\ \bibinfo {pages} {067602} (\bibinfo {year}
  {2010})}\BibitemShut {NoStop}%
\bibitem [{\citenamefont {Pica}\ \emph {et~al.}(2014)\citenamefont {Pica},
  \citenamefont {Wolfowicz}, \citenamefont {Urdampilleta}, \citenamefont
  {Thewalt}, \citenamefont {Riemann}, \citenamefont {Abrosimov}, \citenamefont
  {Becker}, \citenamefont {Pohl}, \citenamefont {Morton}, \citenamefont
  {Bhatt}, \citenamefont {Lyon},\ and\ \citenamefont
  {Lovett}}]{pica_hyperfine_2014}%
  \BibitemOpen
  \bibfield  {author} {\bibinfo {author} {\bibfnamefont {G.}~\bibnamefont
  {Pica}}, \bibinfo {author} {\bibfnamefont {G.}~\bibnamefont {Wolfowicz}},
  \bibinfo {author} {\bibfnamefont {M.}~\bibnamefont {Urdampilleta}}, \bibinfo
  {author} {\bibfnamefont {M.~L.~W.}\ \bibnamefont {Thewalt}}, \bibinfo
  {author} {\bibfnamefont {H.}~\bibnamefont {Riemann}}, \bibinfo {author}
  {\bibfnamefont {N.~V.}\ \bibnamefont {Abrosimov}}, \bibinfo {author}
  {\bibfnamefont {P.}~\bibnamefont {Becker}}, \bibinfo {author} {\bibfnamefont
  {H.-J.}\ \bibnamefont {Pohl}}, \bibinfo {author} {\bibfnamefont {J.~J.~L.}\
  \bibnamefont {Morton}}, \bibinfo {author} {\bibfnamefont {R.~N.}\
  \bibnamefont {Bhatt}}, \bibinfo {author} {\bibfnamefont {S.~A.}\ \bibnamefont
  {Lyon}}, \ and\ \bibinfo {author} {\bibfnamefont {B.~W.}\ \bibnamefont
  {Lovett}},\ }\href {\doibase 10.1103/PhysRevB.90.195204} {\bibfield
  {journal} {\bibinfo  {journal} {Physical Review B}\ }\textbf {\bibinfo
  {volume} {90}},\ \bibinfo {pages} {195204} (\bibinfo {year} {2014})},\
  \bibinfo {note} {publisher: American Physical Society}\BibitemShut {NoStop}%
\bibitem [{\citenamefont {Probst}\ \emph {et~al.}(2019)\citenamefont {Probst},
  \citenamefont {Ranjan}, \citenamefont {Ansel}, \citenamefont {Heeres},
  \citenamefont {Albanese}, \citenamefont {Albertinale}, \citenamefont {Vion},
  \citenamefont {Esteve}, \citenamefont {Glaser}, \citenamefont {Sugny},\ and\
  \citenamefont {Bertet}}]{probst_shaped_2019}%
  \BibitemOpen
  \bibfield  {author} {\bibinfo {author} {\bibfnamefont {S.}~\bibnamefont
  {Probst}}, \bibinfo {author} {\bibfnamefont {V.}~\bibnamefont {Ranjan}},
  \bibinfo {author} {\bibfnamefont {Q.}~\bibnamefont {Ansel}}, \bibinfo
  {author} {\bibfnamefont {R.}~\bibnamefont {Heeres}}, \bibinfo {author}
  {\bibfnamefont {B.}~\bibnamefont {Albanese}}, \bibinfo {author}
  {\bibfnamefont {E.}~\bibnamefont {Albertinale}}, \bibinfo {author}
  {\bibfnamefont {D.}~\bibnamefont {Vion}}, \bibinfo {author} {\bibfnamefont
  {D.}~\bibnamefont {Esteve}}, \bibinfo {author} {\bibfnamefont {S.~J.}\
  \bibnamefont {Glaser}}, \bibinfo {author} {\bibfnamefont {D.}~\bibnamefont
  {Sugny}}, \ and\ \bibinfo {author} {\bibfnamefont {P.}~\bibnamefont
  {Bertet}},\ }\href {\doibase 10.1016/j.jmr.2019.04.008} {\bibfield  {journal}
  {\bibinfo  {journal} {Journal of Magnetic Resonance}\ }\textbf {\bibinfo
  {volume} {303}},\ \bibinfo {pages} {42} (\bibinfo {year} {2019})}\BibitemShut
  {NoStop}%
\bibitem [{\citenamefont {Duzer}\ and\ \citenamefont {{Charles W.
  Turner}}(1999)}]{duzer_principles_1999}%
  \BibitemOpen
  \bibfield  {author} {\bibinfo {author} {\bibfnamefont {T.~V.}\ \bibnamefont
  {Duzer}}\ and\ \bibinfo {author} {\bibnamefont {{Charles W. Turner}}},\
  }\href@noop {} {\emph {\bibinfo {title} {Principles of {Superconductive}
  {Devices} and {Circuits}}}}\ (\bibinfo  {publisher} {Prentice Hall PTR},\
  \bibinfo {year} {1999})\BibitemShut {NoStop}%
\bibitem [{\citenamefont {Purcell}(1946)}]{purcell_spontaneous_1946}%
  \BibitemOpen
  \bibfield  {author} {\bibinfo {author} {\bibfnamefont {E.~M.}\ \bibnamefont
  {Purcell}},\ }\href@noop {} {\bibfield  {journal} {\bibinfo  {journal} {Phys.
  Rev.}\ }\textbf {\bibinfo {volume} {69}},\ \bibinfo {pages} {681} (\bibinfo
  {year} {1946})}\BibitemShut {NoStop}%
\bibitem [{\citenamefont {Bienfait}\ \emph
  {et~al.}(2016{\natexlab{b}})\citenamefont {Bienfait}, \citenamefont {Pla},
  \citenamefont {Kubo}, \citenamefont {Zhou}, \citenamefont {Stern},
  \citenamefont {Lo}, \citenamefont {Weis}, \citenamefont {Schenkel},
  \citenamefont {Vion}, \citenamefont {Esteve}, \citenamefont {Morton},\ and\
  \citenamefont {Bertet}}]{bienfait_controlling_2016}%
  \BibitemOpen
  \bibfield  {author} {\bibinfo {author} {\bibfnamefont {A.}~\bibnamefont
  {Bienfait}}, \bibinfo {author} {\bibfnamefont {J.}~\bibnamefont {Pla}},
  \bibinfo {author} {\bibfnamefont {Y.}~\bibnamefont {Kubo}}, \bibinfo {author}
  {\bibfnamefont {X.}~\bibnamefont {Zhou}}, \bibinfo {author} {\bibfnamefont
  {M.}~\bibnamefont {Stern}}, \bibinfo {author} {\bibfnamefont {C.-C.}\
  \bibnamefont {Lo}}, \bibinfo {author} {\bibfnamefont {C.}~\bibnamefont
  {Weis}}, \bibinfo {author} {\bibfnamefont {T.}~\bibnamefont {Schenkel}},
  \bibinfo {author} {\bibfnamefont {D.}~\bibnamefont {Vion}}, \bibinfo {author}
  {\bibfnamefont {D.}~\bibnamefont {Esteve}}, \bibinfo {author} {\bibfnamefont
  {J.}~\bibnamefont {Morton}}, \ and\ \bibinfo {author} {\bibfnamefont
  {P.}~\bibnamefont {Bertet}},\ }\href {\doibase doi:10.1038/nature16944}
  {\bibfield  {journal} {\bibinfo  {journal} {Nature}\ }\textbf {\bibinfo
  {volume} {531}},\ \bibinfo {pages} {74 } (\bibinfo {year}
  {2016}{\natexlab{b}})}\BibitemShut {NoStop}%
\bibitem [{\citenamefont {Eichler}\ \emph {et~al.}(2017)\citenamefont
  {Eichler}, \citenamefont {Sigillito}, \citenamefont {Lyon},\ and\
  \citenamefont {Petta}}]{eichler_electron_2017}%
  \BibitemOpen
  \bibfield  {author} {\bibinfo {author} {\bibfnamefont {C.}~\bibnamefont
  {Eichler}}, \bibinfo {author} {\bibfnamefont {A.~J.}\ \bibnamefont
  {Sigillito}}, \bibinfo {author} {\bibfnamefont {S.~A.}\ \bibnamefont {Lyon}},
  \ and\ \bibinfo {author} {\bibfnamefont {J.~R.}\ \bibnamefont {Petta}},\
  }\href {\doibase 10.1103/PhysRevLett.118.037701} {\bibfield  {journal}
  {\bibinfo  {journal} {Physical Review Letters}\ }\textbf {\bibinfo {volume}
  {118}},\ \bibinfo {pages} {037701} (\bibinfo {year} {2017})}\BibitemShut
  {NoStop}%
\bibitem [{\citenamefont {Macklin}\ \emph {et~al.}(2015)\citenamefont
  {Macklin}, \citenamefont {O{\textquoteright}Brien}, \citenamefont {Hover},
  \citenamefont {Schwartz}, \citenamefont {Bolkhovsky}, \citenamefont {Zhang},
  \citenamefont {Oliver},\ and\ \citenamefont
  {Siddiqi}}]{macklin_nearquantum-limited_2015}%
  \BibitemOpen
  \bibfield  {author} {\bibinfo {author} {\bibfnamefont {C.}~\bibnamefont
  {Macklin}}, \bibinfo {author} {\bibfnamefont {K.}~\bibnamefont
  {O{\textquoteright}Brien}}, \bibinfo {author} {\bibfnamefont
  {D.}~\bibnamefont {Hover}}, \bibinfo {author} {\bibfnamefont {M.~E.}\
  \bibnamefont {Schwartz}}, \bibinfo {author} {\bibfnamefont {V.}~\bibnamefont
  {Bolkhovsky}}, \bibinfo {author} {\bibfnamefont {X.}~\bibnamefont {Zhang}},
  \bibinfo {author} {\bibfnamefont {W.~D.}\ \bibnamefont {Oliver}}, \ and\
  \bibinfo {author} {\bibfnamefont {I.}~\bibnamefont {Siddiqi}},\ }\href
  {\doibase 10.1126/science.aaa8525} {\bibfield  {journal} {\bibinfo  {journal}
  {Science}\ }\textbf {\bibinfo {volume} {350}},\ \bibinfo {pages} {307}
  (\bibinfo {year} {2015})}\BibitemShut {NoStop}%
\bibitem [{\citenamefont {Ranjan}\ \emph
  {et~al.}(2020{\natexlab{c}})\citenamefont {Ranjan}, \citenamefont {Probst},
  \citenamefont {Albanese}, \citenamefont {Doll}, \citenamefont {Jacquot},
  \citenamefont {Flurin}, \citenamefont {Heeres}, \citenamefont {Vion},
  \citenamefont {Esteve}, \citenamefont {Morton},\ and\ \citenamefont
  {Bertet}}]{ranjan_pulsed_2020}%
  \BibitemOpen
  \bibfield  {author} {\bibinfo {author} {\bibfnamefont {V.}~\bibnamefont
  {Ranjan}}, \bibinfo {author} {\bibfnamefont {S.}~\bibnamefont {Probst}},
  \bibinfo {author} {\bibfnamefont {B.}~\bibnamefont {Albanese}}, \bibinfo
  {author} {\bibfnamefont {A.}~\bibnamefont {Doll}}, \bibinfo {author}
  {\bibfnamefont {O.}~\bibnamefont {Jacquot}}, \bibinfo {author} {\bibfnamefont
  {E.}~\bibnamefont {Flurin}}, \bibinfo {author} {\bibfnamefont
  {R.}~\bibnamefont {Heeres}}, \bibinfo {author} {\bibfnamefont
  {D.}~\bibnamefont {Vion}}, \bibinfo {author} {\bibfnamefont {D.}~\bibnamefont
  {Esteve}}, \bibinfo {author} {\bibfnamefont {J.~J.~L.}\ \bibnamefont
  {Morton}}, \ and\ \bibinfo {author} {\bibfnamefont {P.}~\bibnamefont
  {Bertet}},\ }\href {\doibase https://doi.org/10.1016/j.jmr.2019.106662}
  {\bibfield  {journal} {\bibinfo  {journal} {Journal of Magnetic Resonance}\
  }\textbf {\bibinfo {volume} {310}},\ \bibinfo {pages} {106662} (\bibinfo
  {year} {2020}{\natexlab{c}})}\BibitemShut {NoStop}%
\bibitem [{\citenamefont {Weis}\ \emph {et~al.}(2012)\citenamefont {Weis},
  \citenamefont {Lo}, \citenamefont {Lang}, \citenamefont {Tyryshkin},
  \citenamefont {George}, \citenamefont {Yu}, \citenamefont {Bokor},
  \citenamefont {Lyon}, \citenamefont {Morton},\ and\ \citenamefont
  {Schenkel}}]{weis_electrical_2012}%
  \BibitemOpen
  \bibfield  {author} {\bibinfo {author} {\bibfnamefont {C.~D.}\ \bibnamefont
  {Weis}}, \bibinfo {author} {\bibfnamefont {C.~C.}\ \bibnamefont {Lo}},
  \bibinfo {author} {\bibfnamefont {V.}~\bibnamefont {Lang}}, \bibinfo {author}
  {\bibfnamefont {A.~M.}\ \bibnamefont {Tyryshkin}}, \bibinfo {author}
  {\bibfnamefont {R.~E.}\ \bibnamefont {George}}, \bibinfo {author}
  {\bibfnamefont {K.~M.}\ \bibnamefont {Yu}}, \bibinfo {author} {\bibfnamefont
  {J.}~\bibnamefont {Bokor}}, \bibinfo {author} {\bibfnamefont {S.~A.}\
  \bibnamefont {Lyon}}, \bibinfo {author} {\bibfnamefont {J.~J.~L.}\
  \bibnamefont {Morton}}, \ and\ \bibinfo {author} {\bibfnamefont
  {T.}~\bibnamefont {Schenkel}},\ }\href {\doibase 10.1063/1.4704561}
  {\bibfield  {journal} {\bibinfo  {journal} {Applied Physics Letters}\
  }\textbf {\bibinfo {volume} {100}},\ \bibinfo {pages} {172104} (\bibinfo
  {year} {2012})}\BibitemShut {NoStop}%
\bibitem [{\citenamefont {Sekiguchi}\ \emph {et~al.}(2010)\citenamefont
  {Sekiguchi}, \citenamefont {Steger}, \citenamefont {Saeedi}, \citenamefont
  {Thewalt}, \citenamefont {Riemann}, \citenamefont {Abrosimov},\ and\
  \citenamefont {N{\"o}tzel}}]{sekiguchi_hyperfine_2010}%
  \BibitemOpen
  \bibfield  {author} {\bibinfo {author} {\bibfnamefont {T.}~\bibnamefont
  {Sekiguchi}}, \bibinfo {author} {\bibfnamefont {M.}~\bibnamefont {Steger}},
  \bibinfo {author} {\bibfnamefont {K.}~\bibnamefont {Saeedi}}, \bibinfo
  {author} {\bibfnamefont {M.~L.~W.}\ \bibnamefont {Thewalt}}, \bibinfo
  {author} {\bibfnamefont {H.}~\bibnamefont {Riemann}}, \bibinfo {author}
  {\bibfnamefont {N.~V.}\ \bibnamefont {Abrosimov}}, \ and\ \bibinfo {author}
  {\bibfnamefont {N.}~\bibnamefont {N{\"o}tzel}},\ }\href {\doibase
  10.1103/PhysRevLett.104.137402} {\bibfield  {journal} {\bibinfo  {journal}
  {Physical Review Letters}\ }\textbf {\bibinfo {volume} {104}},\ \bibinfo
  {pages} {137402} (\bibinfo {year} {2010})}\BibitemShut {NoStop}%
\bibitem [{\citenamefont {Albanese}\ \emph {et~al.}(2020)\citenamefont
  {Albanese}, \citenamefont {Probst}, \citenamefont {Ranjan}, \citenamefont
  {Zollitsch}, \citenamefont {Pechal}, \citenamefont {Wallraff}, \citenamefont
  {Morton}, \citenamefont {Vion}, \citenamefont {Esteve}, \citenamefont
  {Flurin},\ and\ \citenamefont {Bertet}}]{albanese_radiative_2020}%
  \BibitemOpen
  \bibfield  {author} {\bibinfo {author} {\bibfnamefont {B.}~\bibnamefont
  {Albanese}}, \bibinfo {author} {\bibfnamefont {S.}~\bibnamefont {Probst}},
  \bibinfo {author} {\bibfnamefont {V.}~\bibnamefont {Ranjan}}, \bibinfo
  {author} {\bibfnamefont {C.~W.}\ \bibnamefont {Zollitsch}}, \bibinfo {author}
  {\bibfnamefont {M.}~\bibnamefont {Pechal}}, \bibinfo {author} {\bibfnamefont
  {A.}~\bibnamefont {Wallraff}}, \bibinfo {author} {\bibfnamefont {J.~J.~L.}\
  \bibnamefont {Morton}}, \bibinfo {author} {\bibfnamefont {D.}~\bibnamefont
  {Vion}}, \bibinfo {author} {\bibfnamefont {D.}~\bibnamefont {Esteve}},
  \bibinfo {author} {\bibfnamefont {E.}~\bibnamefont {Flurin}}, \ and\ \bibinfo
  {author} {\bibfnamefont {P.}~\bibnamefont {Bertet}},\ }\href {\doibase
  10.1038/s41567-020-0872-2} {\bibfield  {journal} {\bibinfo  {journal} {Nature
  Physics}\ }\textbf {\bibinfo {volume} {16}},\ \bibinfo {pages} {751}
  (\bibinfo {year} {2020})}\BibitemShut {NoStop}%
\bibitem [{\citenamefont {Schweiger}\ and\ \citenamefont
  {Jeschke}(2001)}]{schweiger_principles_2001}%
  \BibitemOpen
  \bibfield  {author} {\bibinfo {author} {\bibfnamefont {A.}~\bibnamefont
  {Schweiger}}\ and\ \bibinfo {author} {\bibfnamefont {G.}~\bibnamefont
  {Jeschke}},\ }\href@noop {} {\emph {\bibinfo {title} {Principles of pulse
  electron paramagnetic resonance}}}\ (\bibinfo  {publisher} {Oxford University
  Press},\ \bibinfo {year} {2001})\BibitemShut {NoStop}%
\bibitem [{\citenamefont {Ithier}\ \emph {et~al.}(2005)\citenamefont {Ithier},
  \citenamefont {Collin}, \citenamefont {Joyez}, \citenamefont {Meeson},
  \citenamefont {Vion}, \citenamefont {Esteve}, \citenamefont {Chiarello},
  \citenamefont {Shnirman}, \citenamefont {Makhlin}, \citenamefont {Schriefl},\
  and\ \citenamefont {Sch?n}}]{ithier_decoherence_2005}%
  \BibitemOpen
  \bibfield  {author} {\bibinfo {author} {\bibfnamefont {G.}~\bibnamefont
  {Ithier}}, \bibinfo {author} {\bibfnamefont {E.}~\bibnamefont {Collin}},
  \bibinfo {author} {\bibfnamefont {P.}~\bibnamefont {Joyez}}, \bibinfo
  {author} {\bibfnamefont {P.~J.}\ \bibnamefont {Meeson}}, \bibinfo {author}
  {\bibfnamefont {D.}~\bibnamefont {Vion}}, \bibinfo {author} {\bibfnamefont
  {D.}~\bibnamefont {Esteve}}, \bibinfo {author} {\bibfnamefont
  {F.}~\bibnamefont {Chiarello}}, \bibinfo {author} {\bibfnamefont
  {A.}~\bibnamefont {Shnirman}}, \bibinfo {author} {\bibfnamefont
  {Y.}~\bibnamefont {Makhlin}}, \bibinfo {author} {\bibfnamefont
  {J.}~\bibnamefont {Schriefl}}, \ and\ \bibinfo {author} {\bibfnamefont
  {G.}~\bibnamefont {Sch?n}},\ }\href {\doibase 10.1103/PhysRevB.72.134519}
  {\bibfield  {journal} {\bibinfo  {journal} {Physical Review B}\ }\textbf
  {\bibinfo {volume} {72}},\ \bibinfo {pages} {134519} (\bibinfo {year}
  {2005})},\ \bibinfo {note} {copyright (C) 2009 The American Physical Society;
  Please report any problems to prola@aps.org}\BibitemShut {NoStop}%
\bibitem [{\citenamefont {Yoneda}\ \emph {et~al.}(2018)\citenamefont {Yoneda},
  \citenamefont {Takeda}, \citenamefont {Otsuka}, \citenamefont {Nakajima},
  \citenamefont {Delbecq}, \citenamefont {Allison}, \citenamefont {Honda},
  \citenamefont {Kodera}, \citenamefont {Oda}, \citenamefont {Hoshi},
  \citenamefont {Usami}, \citenamefont {Itoh},\ and\ \citenamefont
  {Tarucha}}]{yoneda_quantum-dot_2018}%
  \BibitemOpen
  \bibfield  {author} {\bibinfo {author} {\bibfnamefont {J.}~\bibnamefont
  {Yoneda}}, \bibinfo {author} {\bibfnamefont {K.}~\bibnamefont {Takeda}},
  \bibinfo {author} {\bibfnamefont {T.}~\bibnamefont {Otsuka}}, \bibinfo
  {author} {\bibfnamefont {T.}~\bibnamefont {Nakajima}}, \bibinfo {author}
  {\bibfnamefont {M.~R.}\ \bibnamefont {Delbecq}}, \bibinfo {author}
  {\bibfnamefont {G.}~\bibnamefont {Allison}}, \bibinfo {author} {\bibfnamefont
  {T.}~\bibnamefont {Honda}}, \bibinfo {author} {\bibfnamefont
  {T.}~\bibnamefont {Kodera}}, \bibinfo {author} {\bibfnamefont
  {S.}~\bibnamefont {Oda}}, \bibinfo {author} {\bibfnamefont {Y.}~\bibnamefont
  {Hoshi}}, \bibinfo {author} {\bibfnamefont {N.}~\bibnamefont {Usami}},
  \bibinfo {author} {\bibfnamefont {K.~M.}\ \bibnamefont {Itoh}}, \ and\
  \bibinfo {author} {\bibfnamefont {S.}~\bibnamefont {Tarucha}},\ }\href
  {\doibase 10.1038/s41565-017-0014-x} {\bibfield  {journal} {\bibinfo
  {journal} {Nature Nanotechnology}\ }\textbf {\bibinfo {volume} {13}},\
  \bibinfo {pages} {102} (\bibinfo {year} {2018})}\BibitemShut {NoStop}%
\bibitem [{\citenamefont {Van~Gorp}\ and\ \citenamefont
  {Stesmans}(1992)}]{van_gorp_dipolar_1992}%
  \BibitemOpen
  \bibfield  {author} {\bibinfo {author} {\bibfnamefont {G.}~\bibnamefont
  {Van~Gorp}}\ and\ \bibinfo {author} {\bibfnamefont {A.}~\bibnamefont
  {Stesmans}},\ }\href {\doibase 10.1103/PhysRevB.45.4344} {\bibfield
  {journal} {\bibinfo  {journal} {Physical Review B}\ }\textbf {\bibinfo
  {volume} {45}},\ \bibinfo {pages} {4344} (\bibinfo {year} {1992})},\ \bibinfo
  {note} {publisher: American Physical Society}\BibitemShut {NoStop}%
\bibitem [{\citenamefont {Koch}\ \emph {et~al.}(2007)\citenamefont {Koch},
  \citenamefont {DiVincenzo},\ and\ \citenamefont {Clarke}}]{koch_model_2007}%
  \BibitemOpen
  \bibfield  {author} {\bibinfo {author} {\bibfnamefont {R.~H.}\ \bibnamefont
  {Koch}}, \bibinfo {author} {\bibfnamefont {D.~P.}\ \bibnamefont
  {DiVincenzo}}, \ and\ \bibinfo {author} {\bibfnamefont {J.}~\bibnamefont
  {Clarke}},\ }\href {\doibase 10.1103/PhysRevLett.98.267003} {\bibfield
  {journal} {\bibinfo  {journal} {Physical Review Letters}\ }\textbf {\bibinfo
  {volume} {98}},\ \bibinfo {pages} {267003} (\bibinfo {year}
  {2007})}\BibitemShut {NoStop}%
\bibitem [{\citenamefont {Sendelbach}\ \emph {et~al.}(2008)\citenamefont
  {Sendelbach}, \citenamefont {Hover}, \citenamefont {Kittel}, \citenamefont
  {M{\"u}ck}, \citenamefont {Martinis},\ and\ \citenamefont
  {McDermott}}]{sendelbach_magnetism_2008}%
  \BibitemOpen
  \bibfield  {author} {\bibinfo {author} {\bibfnamefont {S.}~\bibnamefont
  {Sendelbach}}, \bibinfo {author} {\bibfnamefont {D.}~\bibnamefont {Hover}},
  \bibinfo {author} {\bibfnamefont {A.}~\bibnamefont {Kittel}}, \bibinfo
  {author} {\bibfnamefont {M.}~\bibnamefont {M{\"u}ck}}, \bibinfo {author}
  {\bibfnamefont {J.~M.}\ \bibnamefont {Martinis}}, \ and\ \bibinfo {author}
  {\bibfnamefont {R.}~\bibnamefont {McDermott}},\ }\href {\doibase
  10.1103/PhysRevLett.100.227006} {\bibfield  {journal} {\bibinfo  {journal}
  {Physical Review Letters}\ }\textbf {\bibinfo {volume} {100}},\ \bibinfo
  {pages} {227006} (\bibinfo {year} {2008})}\BibitemShut {NoStop}%
\bibitem [{\citenamefont {Braum{\"u}ller}\ \emph {et~al.}(2020)\citenamefont
  {Braum{\"u}ller}, \citenamefont {Ding}, \citenamefont {Veps{\"a}l{\"a}inen},
  \citenamefont {Sung}, \citenamefont {Kjaergaard}, \citenamefont {Menke},
  \citenamefont {Winik}, \citenamefont {Kim}, \citenamefont {Niedzielski},
  \citenamefont {Melville}, \citenamefont {Yoder}, \citenamefont
  {Hirjibehedin}, \citenamefont {Orlando}, \citenamefont {Gustavsson},\ and\
  \citenamefont {Oliver}}]{braumuller_characterizing_2020}%
  \BibitemOpen
  \bibfield  {author} {\bibinfo {author} {\bibfnamefont {J.}~\bibnamefont
  {Braum{\"u}ller}}, \bibinfo {author} {\bibfnamefont {L.}~\bibnamefont
  {Ding}}, \bibinfo {author} {\bibfnamefont {A.~P.}\ \bibnamefont
  {Veps{\"a}l{\"a}inen}}, \bibinfo {author} {\bibfnamefont {Y.}~\bibnamefont
  {Sung}}, \bibinfo {author} {\bibfnamefont {M.}~\bibnamefont {Kjaergaard}},
  \bibinfo {author} {\bibfnamefont {T.}~\bibnamefont {Menke}}, \bibinfo
  {author} {\bibfnamefont {R.}~\bibnamefont {Winik}}, \bibinfo {author}
  {\bibfnamefont {D.}~\bibnamefont {Kim}}, \bibinfo {author} {\bibfnamefont
  {B.~M.}\ \bibnamefont {Niedzielski}}, \bibinfo {author} {\bibfnamefont
  {A.}~\bibnamefont {Melville}}, \bibinfo {author} {\bibfnamefont {J.~L.}\
  \bibnamefont {Yoder}}, \bibinfo {author} {\bibfnamefont {C.~F.}\ \bibnamefont
  {Hirjibehedin}}, \bibinfo {author} {\bibfnamefont {T.~P.}\ \bibnamefont
  {Orlando}}, \bibinfo {author} {\bibfnamefont {S.}~\bibnamefont {Gustavsson}},
  \ and\ \bibinfo {author} {\bibfnamefont {W.~D.}\ \bibnamefont {Oliver}},\
  }\href {\doibase 10.1103/PhysRevApplied.13.054079} {\bibfield  {journal}
  {\bibinfo  {journal} {Physical Review Applied}\ }\textbf {\bibinfo {volume}
  {13}},\ \bibinfo {pages} {054079} (\bibinfo {year} {2020})}\BibitemShut
  {NoStop}%
\bibitem [{\citenamefont {Lim}\ \emph {et~al.}(2009)\citenamefont {Lim},
  \citenamefont {Huebl}, \citenamefont {Willems~van Beveren}, \citenamefont
  {Rubanov}, \citenamefont {Spizzirri}, \citenamefont {Angus}, \citenamefont
  {Clark},\ and\ \citenamefont {Dzurak}}]{lim_electrostatically_2009}%
  \BibitemOpen
  \bibfield  {author} {\bibinfo {author} {\bibfnamefont {W.~H.}\ \bibnamefont
  {Lim}}, \bibinfo {author} {\bibfnamefont {H.}~\bibnamefont {Huebl}}, \bibinfo
  {author} {\bibfnamefont {L.~H.}\ \bibnamefont {Willems~van Beveren}},
  \bibinfo {author} {\bibfnamefont {S.}~\bibnamefont {Rubanov}}, \bibinfo
  {author} {\bibfnamefont {P.~G.}\ \bibnamefont {Spizzirri}}, \bibinfo {author}
  {\bibfnamefont {S.~J.}\ \bibnamefont {Angus}}, \bibinfo {author}
  {\bibfnamefont {R.~G.}\ \bibnamefont {Clark}}, \ and\ \bibinfo {author}
  {\bibfnamefont {A.~S.}\ \bibnamefont {Dzurak}},\ }\href {\doibase
  10.1063/1.3124242} {\bibfield  {journal} {\bibinfo  {journal} {Applied
  Physics Letters}\ }\textbf {\bibinfo {volume} {94}},\ \bibinfo {pages}
  {173502} (\bibinfo {year} {2009})}\BibitemShut {NoStop}%
\bibitem [{\citenamefont {Spruijtenburg}\ \emph {et~al.}(2018)\citenamefont
  {Spruijtenburg}, \citenamefont {Amitonov}, \citenamefont {Wiel},\ and\
  \citenamefont {Zwanenburg}}]{spruijtenburg_fabrication_2018}%
  \BibitemOpen
  \bibfield  {author} {\bibinfo {author} {\bibfnamefont {P.~C.}\ \bibnamefont
  {Spruijtenburg}}, \bibinfo {author} {\bibfnamefont {S.~V.}\ \bibnamefont
  {Amitonov}}, \bibinfo {author} {\bibfnamefont {W.~G. v.~d.}\ \bibnamefont
  {Wiel}}, \ and\ \bibinfo {author} {\bibfnamefont {F.~A.}\ \bibnamefont
  {Zwanenburg}},\ }\href {\doibase 10.1088/1361-6528/aaabf5} {\bibfield
  {journal} {\bibinfo  {journal} {Nanotechnology}\ }\textbf {\bibinfo {volume}
  {29}},\ \bibinfo {pages} {143001} (\bibinfo {year} {2018})}\BibitemShut
  {NoStop}%
\bibitem [{\citenamefont {Petit}\ \emph {et~al.}(2018)\citenamefont {Petit},
  \citenamefont {Boter}, \citenamefont {Eenink}, \citenamefont {Droulers},
  \citenamefont {Tagliaferri}, \citenamefont {Li}, \citenamefont {Franke},
  \citenamefont {Singh}, \citenamefont {Clarke}, \citenamefont {Schouten},
  \citenamefont {Dobrovitski}, \citenamefont {Vandersypen},\ and\ \citenamefont
  {Veldhorst}}]{petit_spin_2018}%
  \BibitemOpen
  \bibfield  {author} {\bibinfo {author} {\bibfnamefont {L.}~\bibnamefont
  {Petit}}, \bibinfo {author} {\bibfnamefont {J.~M.}\ \bibnamefont {Boter}},
  \bibinfo {author} {\bibfnamefont {H.~G.~J.}\ \bibnamefont {Eenink}}, \bibinfo
  {author} {\bibfnamefont {G.}~\bibnamefont {Droulers}}, \bibinfo {author}
  {\bibfnamefont {M.~L.~V.}\ \bibnamefont {Tagliaferri}}, \bibinfo {author}
  {\bibfnamefont {R.}~\bibnamefont {Li}}, \bibinfo {author} {\bibfnamefont
  {D.~P.}\ \bibnamefont {Franke}}, \bibinfo {author} {\bibfnamefont {K.~J.}\
  \bibnamefont {Singh}}, \bibinfo {author} {\bibfnamefont {J.~S.}\ \bibnamefont
  {Clarke}}, \bibinfo {author} {\bibfnamefont {R.~N.}\ \bibnamefont
  {Schouten}}, \bibinfo {author} {\bibfnamefont {V.~V.}\ \bibnamefont
  {Dobrovitski}}, \bibinfo {author} {\bibfnamefont {L.~M.~K.}\ \bibnamefont
  {Vandersypen}}, \ and\ \bibinfo {author} {\bibfnamefont {M.}~\bibnamefont
  {Veldhorst}},\ }\href {\doibase 10.1103/PhysRevLett.121.076801} {\bibfield
  {journal} {\bibinfo  {journal} {Physical Review Letters}\ }\textbf {\bibinfo
  {volume} {121}},\ \bibinfo {pages} {076801} (\bibinfo {year}
  {2018})}\BibitemShut {NoStop}%
\bibitem [{\citenamefont {Connors}\ \emph {et~al.}(2019)\citenamefont
  {Connors}, \citenamefont {Nelson}, \citenamefont {Qiao}, \citenamefont
  {Edge},\ and\ \citenamefont {Nichol}}]{connors_low-frequency_2019}%
  \BibitemOpen
  \bibfield  {author} {\bibinfo {author} {\bibfnamefont {E.~J.}\ \bibnamefont
  {Connors}}, \bibinfo {author} {\bibfnamefont {J.}~\bibnamefont {Nelson}},
  \bibinfo {author} {\bibfnamefont {H.}~\bibnamefont {Qiao}}, \bibinfo {author}
  {\bibfnamefont {L.~F.}\ \bibnamefont {Edge}}, \ and\ \bibinfo {author}
  {\bibfnamefont {J.~M.}\ \bibnamefont {Nichol}},\ }\href {\doibase
  10.1103/PhysRevB.100.165305} {\bibfield  {journal} {\bibinfo  {journal}
  {Physical Review B}\ }\textbf {\bibinfo {volume} {100}},\ \bibinfo {pages}
  {165305} (\bibinfo {year} {2019})}\BibitemShut {NoStop}%
\bibitem [{\citenamefont {{\'A}lvarez}\ and\ \citenamefont
  {Suter}(2011)}]{alvarez_measuring_2011}%
  \BibitemOpen
  \bibfield  {author} {\bibinfo {author} {\bibfnamefont {G.~A.}\ \bibnamefont
  {{\'A}lvarez}}\ and\ \bibinfo {author} {\bibfnamefont {D.}~\bibnamefont
  {Suter}},\ }\href {\doibase 10.1103/PhysRevLett.107.230501} {\bibfield
  {journal} {\bibinfo  {journal} {Physical Review Letters}\ }\textbf {\bibinfo
  {volume} {107}},\ \bibinfo {pages} {230501} (\bibinfo {year}
  {2011})}\BibitemShut {NoStop}%
\bibitem [{\citenamefont {Yuge}\ \emph {et~al.}(2011)\citenamefont {Yuge},
  \citenamefont {Sasaki},\ and\ \citenamefont
  {Hirayama}}]{yuge_measurement_2011}%
  \BibitemOpen
  \bibfield  {author} {\bibinfo {author} {\bibfnamefont {T.}~\bibnamefont
  {Yuge}}, \bibinfo {author} {\bibfnamefont {S.}~\bibnamefont {Sasaki}}, \ and\
  \bibinfo {author} {\bibfnamefont {Y.}~\bibnamefont {Hirayama}},\ }\href
  {\doibase 10.1103/PhysRevLett.107.170504} {\bibfield  {journal} {\bibinfo
  {journal} {Physical Review Letters}\ }\textbf {\bibinfo {volume} {107}},\
  \bibinfo {pages} {170504} (\bibinfo {year} {2011})}\BibitemShut {NoStop}%
\bibitem [{\citenamefont {Medford}\ \emph {et~al.}(2012)\citenamefont
  {Medford}, \citenamefont {Cywi{\'n}ski}, \citenamefont {Barthel},
  \citenamefont {Marcus}, \citenamefont {Hanson},\ and\ \citenamefont
  {Gossard}}]{medford_scaling_2012}%
  \BibitemOpen
  \bibfield  {author} {\bibinfo {author} {\bibfnamefont {J.}~\bibnamefont
  {Medford}}, \bibinfo {author} {\bibfnamefont {{\L }.}~\bibnamefont
  {Cywi{\'n}ski}}, \bibinfo {author} {\bibfnamefont {C.}~\bibnamefont
  {Barthel}}, \bibinfo {author} {\bibfnamefont {C.~M.}\ \bibnamefont {Marcus}},
  \bibinfo {author} {\bibfnamefont {M.~P.}\ \bibnamefont {Hanson}}, \ and\
  \bibinfo {author} {\bibfnamefont {A.~C.}\ \bibnamefont {Gossard}},\ }\href
  {\doibase 10.1103/PhysRevLett.108.086802} {\bibfield  {journal} {\bibinfo
  {journal} {Physical Review Letters}\ }\textbf {\bibinfo {volume} {108}},\
  \bibinfo {pages} {086802} (\bibinfo {year} {2012})}\BibitemShut {NoStop}%
\bibitem [{\citenamefont {Myers}\ \emph {et~al.}(2014)\citenamefont {Myers},
  \citenamefont {Das}, \citenamefont {Dartiailh}, \citenamefont {Ohno},
  \citenamefont {Awschalom},\ and\ \citenamefont
  {Bleszynski~Jayich}}]{myers_probing_2014}%
  \BibitemOpen
  \bibfield  {author} {\bibinfo {author} {\bibfnamefont {B.~A.}\ \bibnamefont
  {Myers}}, \bibinfo {author} {\bibfnamefont {A.}~\bibnamefont {Das}}, \bibinfo
  {author} {\bibfnamefont {M.~C.}\ \bibnamefont {Dartiailh}}, \bibinfo {author}
  {\bibfnamefont {K.}~\bibnamefont {Ohno}}, \bibinfo {author} {\bibfnamefont
  {D.~D.}\ \bibnamefont {Awschalom}}, \ and\ \bibinfo {author} {\bibfnamefont
  {A.~C.}\ \bibnamefont {Bleszynski~Jayich}},\ }\href {\doibase
  10.1103/PhysRevLett.113.027602} {\bibfield  {journal} {\bibinfo  {journal}
  {Physical Review Letters}\ }\textbf {\bibinfo {volume} {113}},\ \bibinfo
  {pages} {027602} (\bibinfo {year} {2014})}\BibitemShut {NoStop}%
\bibitem [{\citenamefont {Yoshihara}\ \emph {et~al.}(2006)\citenamefont
  {Yoshihara}, \citenamefont {Harrabi}, \citenamefont {Niskanen}, \citenamefont
  {Nakamura},\ and\ \citenamefont {Tsai}}]{yoshihara_decoherence_2006}%
  \BibitemOpen
  \bibfield  {author} {\bibinfo {author} {\bibfnamefont {F.}~\bibnamefont
  {Yoshihara}}, \bibinfo {author} {\bibfnamefont {K.}~\bibnamefont {Harrabi}},
  \bibinfo {author} {\bibfnamefont {A.~O.}\ \bibnamefont {Niskanen}}, \bibinfo
  {author} {\bibfnamefont {Y.}~\bibnamefont {Nakamura}}, \ and\ \bibinfo
  {author} {\bibfnamefont {J.~S.}\ \bibnamefont {Tsai}},\ }\href {\doibase
  10.1103/PhysRevLett.97.167001} {\bibfield  {journal} {\bibinfo  {journal}
  {Phys. Rev. Lett.}\ }\textbf {\bibinfo {volume} {97}},\ \bibinfo {pages}
  {167001} (\bibinfo {year} {2006})}\BibitemShut {NoStop}%
\bibitem [{\citenamefont {Bylander}\ \emph {et~al.}(2011)\citenamefont
  {Bylander}, \citenamefont {Gustavsson}, \citenamefont {Yan}, \citenamefont
  {Yoshihara}, \citenamefont {Harrabi}, \citenamefont {Fitch}, \citenamefont
  {Cory}, \citenamefont {Nakamura}, \citenamefont {Tsai},\ and\ \citenamefont
  {Oliver}}]{bylander_noise_2011}%
  \BibitemOpen
  \bibfield  {author} {\bibinfo {author} {\bibfnamefont {J.}~\bibnamefont
  {Bylander}}, \bibinfo {author} {\bibfnamefont {S.}~\bibnamefont
  {Gustavsson}}, \bibinfo {author} {\bibfnamefont {F.}~\bibnamefont {Yan}},
  \bibinfo {author} {\bibfnamefont {F.}~\bibnamefont {Yoshihara}}, \bibinfo
  {author} {\bibfnamefont {K.}~\bibnamefont {Harrabi}}, \bibinfo {author}
  {\bibfnamefont {G.}~\bibnamefont {Fitch}}, \bibinfo {author} {\bibfnamefont
  {D.~G.}\ \bibnamefont {Cory}}, \bibinfo {author} {\bibfnamefont
  {Y.}~\bibnamefont {Nakamura}}, \bibinfo {author} {\bibfnamefont {J.-S.}\
  \bibnamefont {Tsai}}, \ and\ \bibinfo {author} {\bibfnamefont {W.~D.}\
  \bibnamefont {Oliver}},\ }\href {\doibase 10.1038/nphys1994} {\bibfield
  {journal} {\bibinfo  {journal} {Nat Phys}\ }\textbf {\bibinfo {volume} {7}},\
  \bibinfo {pages} {565} (\bibinfo {year} {2011})}\BibitemShut {NoStop}%
\bibitem [{\citenamefont {Bertet}\ \emph {et~al.}(2005)\citenamefont {Bertet},
  \citenamefont {Chiorescu}, \citenamefont {Burkard}, \citenamefont {Semba},
  \citenamefont {Harmans}, \citenamefont {DiVincenzo},\ and\ \citenamefont
  {Mooij}}]{bertet_dephasing_2005}%
  \BibitemOpen
  \bibfield  {author} {\bibinfo {author} {\bibfnamefont {P.}~\bibnamefont
  {Bertet}}, \bibinfo {author} {\bibfnamefont {I.}~\bibnamefont {Chiorescu}},
  \bibinfo {author} {\bibfnamefont {G.}~\bibnamefont {Burkard}}, \bibinfo
  {author} {\bibfnamefont {K.}~\bibnamefont {Semba}}, \bibinfo {author}
  {\bibfnamefont {C.~J. P.~M.}\ \bibnamefont {Harmans}}, \bibinfo {author}
  {\bibfnamefont {D.~P.}\ \bibnamefont {DiVincenzo}}, \ and\ \bibinfo {author}
  {\bibfnamefont {J.~E.}\ \bibnamefont {Mooij}},\ }\href {\doibase
  10.1103/PhysRevLett.95.257002} {\bibfield  {journal} {\bibinfo  {journal}
  {Physical Review Letters}\ }\textbf {\bibinfo {volume} {95}},\ \bibinfo
  {pages} {257002} (\bibinfo {year} {2005})}\BibitemShut {NoStop}%
\bibitem [{\citenamefont {Milo{\v s}evi{\'c}}\ \emph
  {et~al.}(2002)\citenamefont {Milo{\v s}evi{\'c}}, \citenamefont
  {Yampolskii},\ and\ \citenamefont {Peeters}}]{milosevic_magnetic_2002}%
  \BibitemOpen
  \bibfield  {author} {\bibinfo {author} {\bibfnamefont {M.~V.}\ \bibnamefont
  {Milo{\v s}evi{\'c}}}, \bibinfo {author} {\bibfnamefont {S.~V.}\ \bibnamefont
  {Yampolskii}}, \ and\ \bibinfo {author} {\bibfnamefont {F.~M.}\ \bibnamefont
  {Peeters}},\ }\href {\doibase 10.1103/PhysRevB.66.174519} {\bibfield
  {journal} {\bibinfo  {journal} {Physical Review B}\ }\textbf {\bibinfo
  {volume} {66}},\ \bibinfo {pages} {174519} (\bibinfo {year} {2002})},\
  \bibinfo {note} {publisher: American Physical Society}\BibitemShut {NoStop}%
\end{thebibliography}
%

\end{document}